\documentclass[reprint,aps,a4paper,showpacs,prc,floatfix,twocolumn,amsmath,amssymb]{revtex4-1}
\usepackage{amsmath,amssymb,amsfonts}

\usepackage{amsmath,amssymb,amsfonts}
\usepackage{graphicx}
\usepackage{xcolor}
\usepackage{floatflt}
\usepackage{longtable}
\usepackage{dcolumn}
\usepackage[normalem]{ulem}

\allowdisplaybreaks

\newcommand{\be}{\begin{equation}}
\newcommand{\ee}{\end{equation}}
\newcommand{\bea}{\begin{eqnarray}}
\newcommand{\eea}{\end{eqnarray}}
\def\msun{{\rm M}_\odot}

\begin{document}



\title{(No) neutron star maximum mass constraint from hypernuclei}

\author {M. Fortin} 
\affiliation{N. Copernicus Astronomical Center, Polish Academy of Sciences,
Bartycka 18, 00-716 Warszawa, Poland}
\author{S. S. Avancini} 
\affiliation{Depto de F\'{\i}sica - CFM - Universidade Federal de Santa
Catarina  Florian\'opolis - SC - CP. 476 - CEP 88.040 - 900, Brazil}
\author{C. Provid{\^e}ncia} 
\author{I. Vida\~na}
\affiliation{CFisUC, Department of Physics, University of Coimbra, 3004-516 Coimbra, Portugal}


\begin{abstract}
\begin{description}
\item[Background] 

 The recently accurate measurement of the mass of two pulsars
  close to or above $2\, M_\odot$ has raised the question whether such
  large pulsar masses allow for the existence of exotic degrees of
  freedom, such as hyperons, inside neutron stars.

\item[Purpose] 

 In the present work we will investigate how  the existing hypernuclei
properties may constrain the neutron star equation of state and
confront the neutron star maximum masses obtained with equations of state
calibrated to hypernuclei properties with the astrophysical $2\,M_\odot$ constraint. 

\item[Method] 

The study is performed using a relativistic mean field approach to
  describe both the hypernuclei and the neutron star equations of
  state. Unified equations of state are obtained. A set of five
  models that describe $2\,M_\odot$ when only nucleonic degrees of
  freedom are employed. Some of these models also satisfy
  other well
established laboratory or theoretical constraints.

\item[Results]

The $\Lambda$-meson couplings are determined for all the models
considered, and  the $\Lambda$ potential in symmetric nuclear matter and
$\Lambda$ matter at saturation are calculated. Maximum neutron star
masses are determined for two values of the $\Lambda-\omega$ meson coupling,
$g_{\omega \Lambda}=2g_{\omega N}/3$ and $g_{\omega \Lambda}=g_{\omega N}$,  and a wide range of values for
$g_{\phi \Lambda}$. Hyperonic stars with
the complete baryonic octet are studied, restricting the coupling of the
$\Sigma$ and $\Xi$ hyperons to the $\omega-$, $\rho-$ and $\sigma-$mesons due to
the lack of experimental data, and maximum star masses calculated.

\item[Conclusions]

 We conclude that the currently available hypernuclei experimental data and the
  lack of constraints on the asymmetric equation of state of nuclear
  matter at high densities do not allow to further constrain the neutron star
  matter equation of state using the recent $2\, M_\odot$ observations. It is shown that the  $\Lambda$ potential in symmetric
  nuclear matter takes a value $\sim 30-32$ MeV at saturation for the $g_{\omega \Lambda}$ coupling
  given by the SU(6) symmetry, being of the order of
the values generally used in the literature. On the other hand, the
$\Lambda$ potential in $\Lambda$ matter varies between -16 and -8 MeV
taking for vector mesons couplings the SU(6) values, at variance with generally employed values between $-1$ and $-5$ MeV.
If the SU(6) constraint is relaxed and the vector meson couplings to
hyperons is kept to values not larger than to nucleons then values  between $-17$ and $+9$ MeV are obtained.
\end{description}
\end{abstract}


\maketitle

\section{Introduction}

Neutron stars are among the smallest and densest objects in the Universe. With radii of the order of $\sim 10$ km and masses that can be at least as large as two solar masses, matter inside neutron stars is subject to  
extreme conditions of density, isospin asymmetry and magnetic field intensities. These objects constitute perfect laboratories to study nuclear matter under extreme conditions and the QCD phase diagram at low temperatures and high densities, and, therefore, they have been attracting the attention of different fields of physics. Traditionally neutron star matter has been modelled as a uniform neutron-rich fluid in equilibrium with respect to the weak interactions ($\beta$-stable matter) surrounded by a non-homogeneous crust. Neutrons in the inner crust and neutrons and protons in the uniform core of the star are expected to be superfluid. Due to the large value of the density, new degrees of freedom are expected to appear in the inner core of neutron stars in addition to nucleons. Among others, hyperons, Bose-Einstein condensates of kaons or pions, or even deconfined quark matter have been considered.

Contrary to terrestrial conditions, where hyperons are unstable and decay into nucleons through weak interactions, matter in neutron stars maintains the weak equilibrium between the decays and their inverse capture processes.  Since the pioneering work of Ambartsumyan and Saakyan in 1960 \cite{ambart}, the presence of hyperons in neutron stars has been studied by many authors using either microscopic \cite{micro,vlowk,dbhf1,dbhf2,qmc} or phenomenological \cite{rmf,shf} approaches to the neutron star matter equation of state (EoS). All these works agree that hyperons may appear in the inner core of neutron stars at densities around $\sim (2-3)\times n_0$ ($n_0=0.16$ fm$^{-3}$) when the nucleon chemical potential is large enough to make the conversion of a nucleon into a hyperon energetically favourable. This conversion relieves the Fermi pressure exerted by nucleons making the EoS softer. Consequently the mass of the star, and, in particular, its maximum value $M_{\rm max}$,  is substantially reduced. In microscopic calculations (see {\it e.g.,} Refs.\ \cite{micro,vlowk}), this reduction can be even below the value of the mass of the Hulse--Taylor pulsar ($1.4408 \pm 0.0003\,M_\odot$) \cite{hulsetaylor}. This is not the case, however, in phenomenological calculations which find values of $M_{\rm max}$ compatible with the canonical value above. In fact, most relativistic mean field (RMF) models including hyperons predict maximum masses in the range $1.4-1.8\,M_\odot$ \cite{rmf}, although with some parametrizations masses as large as $2\,M_\odot$ could be even obtained \cite{bombaci08,cavagnoli11}.

The presence of hyperons in neutron stars seems to be energetically unavoidable, although
the strong softening of the EoS associated with their appearance (notably in microscopic models) leads to the prediction of maximum masses not compatible with observations. A natural question, therefore, arises: can hyperons still be present in the interior of neutron stars if $M_{\rm max}$ is reduced to values not compatible with astrophysical observations, although their presence is energetically favourable? This question is at the origin of what has been called the ``hyperon puzzle". Its non-trivial solution is currently a subject of intense research, specially in view of 
the recent measurements of unusually high masses of the millisecond pulsars PSR J1614-2230 ($1.928 \pm 0.017\, M_\odot$) \cite{demorest,fonseca}, and PSR J0348+0432 ($2.01 \pm 0.04\, M_\odot$) \cite{antoniadis} which ruled out almost all currently proposed EoS with hyperons.
The solution of this problem demands a mechanism that could eventually provide the additional repulsion  to 
make the EoS stiffer and the maximum mass compatible with observation \cite{fortin2015}. Three different mechanisms have been proposed: (i) the inclusion of a repulsive hyperon-hyperon interaction through the exchange of vector mesons \cite{Bednarek11,Weissenborn,Oertel14,Maslov}, or less attractive scalar $\sigma$ meson exchange \cite{Dalen}, at the cost of potentially making the EoS too stiff around and below saturation density and, therefore, incompatible with recent quantum Monte Carlo nuclear matter
\cite{Gandolfi12} and chiral effective field theory \cite{Hebeler13} calculations, (ii) the inclusion of repulsive hyperonic three-body forces \cite{taka,vidanatbf,yamamoto,lonardoniprl}, or (iii) the possibility of a phase transition to deconfined quark matter at densities below the hyperon threshold 
\cite{Ozel,WeissenbornSagert,Klahn2013,Bonanno,Lastowiecki2012}. An alternative way to circumvent the hyperon puzzle by invoking the appearance of other hadronic degrees of freedom such as for instance the $\Delta$ isobar that push the onset of hyperons to higher densities has also been  considered \cite{Drago}. We note that very recently, Haidenbauer {\it et al.,} \cite{haidenbauer16} have shown that the $\Lambda$
single-particle potential, obtained in a Brueckner--Hartree--Fock calculation using a hyperon-nucleon interaction derived from an SU(3) chiral effective field theory, becomes strongly repulsive for densities larger than $2n_0$, therefore, shifting the onset of hyperons to extremely high densities potentially solving
the hyperon puzzle without the necessity of invoking any of these more exotic mechanisms.

In addition to the  observation of massive neutron stars, {more} astrophysical constraints on the neutron star EoS, and, consequently, on its hyperon content such as the measurement of their radius, moment of inertia, or the surface gravitational redshift from spectral lines may come in the near future thanks to the next generation of X-ray telescopes or radio observatories. No measurement of the two latter quantities has been obtained so far. Many techniques have been devised to determine neutron star radii but current estimates are still controversial both on theoretical and observational grounds (see {\it e.g.,} discussions in \cite{fortin2015,Miller2016,Haensel2016}. The future X-ray missions such as NICER \cite{nicer}, Athena \cite{athena} and potential LOFT-like missions \cite{loft} promise simultaneous determinations of the mass and radius with a $\sim 5\%$ precision.

In the present work we analyse the possibility of obtaining two solar mass hyperonic stars within the relativistic mean field (RMF) approach  when the hyperon-meson couplings are constrained by the existing experimental hypernuclear data. We shall consider a set of models that satisfy the two solar mass constraint imposed by the pulsars J1614-2230 and  J0348+0432 when only purely nucleonic degrees of freedom are considered, and discuss the consequences of including hyperons when hypernuclear data is used to constrain the hyperon-nucleon and the hyperon-hyperon interactions.  In particular, the experimental data on single and double $\Lambda$-hypernuclei will be taken into account in the model within the framework of the RMF approach as it is done in Ref.\ \cite{Shen06}.  
Recently, a study with a similar objective has been performed in Ref.\ \cite{Sedrakian14}. The authors of this work used symmetry arguments to fix the couplings of the vector mesons to hyperons 
and single $\Lambda$-hypernuclei binding energies to constrain the coupling of the $\sigma$-meson to the $\Lambda$-hyperon. The coupling of the other hyperons to the $\sigma$-meson were  obtained
requiring that the lower bound on the maximum mass of the star is $2\,M_\odot$. In this work, we follow  the same procedure to fix the $\Lambda-\sigma$-meson coupling but a different approach is used 
 for the other couplings. In particular,  we take also into account the experimental data on double $\Lambda$-hypernuclei. A comparison between the two approaches is presented.

 The manuscript is organized in the following way.  In section \ref{sect:RMF} we describe and summarize the main properties of the different RMF parametrizations used in this work. Then,  in section \ref{sect:Yphi}, we review the current status of available hypernuclear experimental data. In section \ref{sect:calib}, we  explain how the binding energies of single and double $\Lambda$-hypernuclei are used to calibrate the coupling constants of the $\Lambda$ hyperon with the different mesons in the RMF models. We confront our results for the $\Lambda$ potential at saturation with the usual values taken in the literature and provide tables with values of the $\Lambda$-couplings calibrated to up-to-date hypernuclear data. Hyperonic and unified EOS are then built in section \ref{sect:NS} and their predictions for $M_{\rm max}$ are confronted to the existence of $2\,\msun$ objects.  We finish by shortly summarizing our results and presenting our conclusions
in section \ref{sect:conc}.

\section{RMF models}
\label{sect:RMF}

\begin{table*}
\begin{ruledtabular}
\center\begin{tabular}{lcccccccc}
Model& $n_0$       & $E_0$ & $K$  & $J$  & $L$ &$K_{\rm sym}$& $M_{\rm max}^{\rm N}$& Ref\\
       & (fm$^{-3}$) &(MeV) & (MeV)& (MeV)&(MeV)& (MeV)  & ($\msun$)     \\
\hline
TM1                & 0.146& -16.3 & 281.2 & 36.9 & 111.2 &  33.8 &  2.18 &\cite{TM1}\\
TM2${\omega \rho}$ & 0.146&- 16.4 & 281.7 & 32.1 &  54.8  &-70.5 & 2.25  &\cite{providencia13} \\
NL3                & 0.149 & -16.2 & 271.6 & 37.4 & 118.9 & 101.6 &2.77  &\cite{NL3}\\
NL3${\omega \rho}$ & 0.148 & -16.2 & 271.6 & 31.7 &  55.5 & -7.6  & 2.75  &\cite{NL3wra}\\
DDME2              & 0.152 & -16.1 & 250.9 & 32.3 &  51.2 & -87.1 & 2.48 &\cite{DDME2}\\
\end{tabular}
\caption{Nuclear properties at saturation density ($n_0$) predicted by the different RMF models used in this work: energy per nucleon ($E_0$), compression modulus ($K$), symmetry energy
($J$), its slope ($L$) and 
incompressibility ($K_{\rm sym}$) at the saturation point of uniform symmetric 
nuclear matter at the density $n_0$. The last column shows the value of neutron star maximum mass 
($M_{\rm max}^{\rm N}$)
predicted by these models when only nucleonic degrees of freedom are considered.}
\label{tab:rmf}
\end{ruledtabular}
\end{table*}

\begin{table}
\begin{ruledtabular}

\centering
\begin{tabular}{cccccccc}
  &  {TM1}  &  {TM2}$\omega\rho$  &  {NL3} &  {NL3$\omega\rho$}   &  {DDME2}  \\ 
\hline
$m_{\sigma}$  & 511.198 & 511.198 & 508.194  & 508.194    & 550.1238 \\ 
$m_{\omega}$ & 783        & 783        & 782.501        & 782.501     & 783 \\ 
$m_{\rho}$     & 770        & 770         & 763        & 763     & 763 \\ 
$g_{\sigma}$            & 10.029   &  9.998     & 10.217   & 10.217  & 10.5396\\ 
$g_{\omega}$           & 12.614   & 12.503    & 12.868   & 12.868 & 13.0189 \\ 
$g_{\rho}$               & 9.264     & 11.303    & 8.948      & 11.277  & 7.3672\\ 
$\kappa/M$              & 3.043      &3.523      & 10.431 & 10.431 & 0 \\ 
$\lambda$             & 3.710    & -47.362 & -28.885 & -28.885 & 0 \\ 
$\xi$                     & 0.0169    & 0.0113 & 0 & 0 & 0 \\ 
$\Lambda_{\rm v}$  & 0             & 0.03  & 0 & 0.03 & 0 \\ 

\end{tabular}
\caption{Parameter sets used in this work.  The DDME2 parameters are
  defined at saturation density and the meson masses are given in MeV.}
\label{tab:parameters}
\end{ruledtabular}

\end{table}

In the following, five different RMF models, all predicting $2\,M_\odot$
purely nucleonic stars, are considered: 
four non-linear Walecka type models with constant coupling parameters and one
density-dependent model with  coupling
parameters that depend on the density.
Among the first, we consider the parametrizations TM1
\cite{TM1}, TM2$\omega\rho$ \cite{providencia13}, NL3 \cite{NL3} and
NL3$\omega\rho$ \cite{NL3wra}, and 
for
the latter we
choose the model DDME2 \cite{DDME2}. Some of their nuclear properties
as well as their prediction for the neutron star maximum mass are presented in Table~\ref{tab:rmf}.
In the following we briefly explain the reasons for the choice of these particular models. 

The parametrization  TM1
\cite{TM1} was used in Ref.\ \cite{Shen06} to describe
single and double $\Lambda$-hypernuclei, and,  we will consider it as a reference. 
This model includes a
non-linear $\omega-$meson term  which  softens the EoS at high densities and  is the
underlying model of the Shen--Toki--Oyamatzu--Sumiyoshi supernova EoS
 \cite{stosa,stosb}. However, this EoS has a too large
symmetry energy slope parameter  ($L=110$ MeV) and does not satisfy the
subsaturation neutron matter  constraints
imposed by microscopic calculations \cite{Hebeler13}.  Including a non-linear term that
mixes the $\omega-$ and $\rho-$mesons allows to overcome these two
shortcomings. This term has been added to the TM1 parametrization,
resulting in the parametrization  TM2$\omega\rho$ \cite{providencia13} that
besides has a weaker non-linear $\omega$ term turning the EoS stiffer
than TM1 at large densities.  We also consider the  NL3 parametrization
\cite{NL3} which was fitted to the ground state properties of both
stable and unstable nuclei. This parametrization predicts very large purely nucleonic
neutron star maximum masses but has the drawback of having, as TM1 , a too large
symmetry energy slope ($L=118$ MeV). Thus we will also consider  the parametrization
NL3$\omega\rho$ \cite{NL3wra} with a softer
density dependence of the symmetry energy 
due to inclusion of the
non-linear $\omega\rho$ term. We note here that in Ref.\ \cite{Fortin16}
this parametrization was one of the few (only four) parametrizations chosen
as satisfying a set of consensual constraints and still able of describing
$2\,M_\odot$ stars. The model DDME2 with density dependent couplings
 was another one of these four parametrizations which we will
also choose in the present study. We note also that of  these five
parametrizations only TM1, TM2$\omega\rho$ and DDME2 satisfy the constraints imposed by the
flow of matter in heavy ion collisions \cite{danielewicz02} (see the discussion in Ref.\ \cite{Dutra14}). However, since the analysis of the experimental flow data is quite complex  and not totally  model independent, this constraint should be taken
with care. Therefore, we will also consider the two parametrizations NL3 and NL3$\omega\rho$.
The set of parameters of all the models is shown in Table \ref{tab:parameters}. The parameters for
the DDME2 model are shown at saturation density.

The inclusion of hyperons in RMF models is performed in a quite natural
way \cite{gm91,Schaffner96}.  The hyperon-nucleon (YN) interaction is described
by means of the exchange of $\sigma-$, $\omega-$ and $\rho$-mesons
similarly to the nucleon-nucleon (NN) one. The hyperon-hyperon (YY) interaction is 
included in our model by considering also the coupling of hyperons with
the hidden strangeness mesons $\sigma^*$ and $\phi$. The
Lagrangian density for a system that includes the eight lightest
baryons, {\it i.e.,} the nucleon doublet (neutron $n$ and proton $p$) and  the six
lightest hyperons ($\Lambda$, the $\Sigma^+,\Sigma^0,\Sigma^-$ triplet, and the 
$\Xi^0, \Xi^-$ doublet), reads \cite{TM1,providencia13}:
\bea
{\cal L}&=&\sum_{B}\bar{\Psi}_{B} \left[\gamma_{\mu}D^{\mu}_{B}- m^{*}_{B}\right]\Psi_{B} \cr
&+& \sum_{l=e,\mu} \bar{\psi}_{l}\left[i\gamma_{\mu}\partial^{\mu}-m_{l}\right]\psi_{l} \cr
&+&\frac{1}{2}\left(\partial_{\mu}\sigma \partial^{\mu}\sigma-m^{2}_{\sigma}\sigma^{2}\right)
-\frac{1}{3!}k\sigma^3-\frac{1}{4!}\lambda\sigma^4 \cr
&+&\frac{1}{2} m^{2}_{\omega}\omega_{\mu}\omega^{\mu}
-\frac{1}{4} \Omega_{\mu \nu} \Omega^{\mu \nu}
+\frac{1}{4!}\xi g_{\omega}^4 \left(\omega_{\mu}\omega^{\mu}\right)^2 \cr
&+&\frac{1}{2}m^{2}_{\rho}\boldsymbol{\rho}_{\mu}\cdot\boldsymbol{\rho}^{\mu}
-\frac{1}{4} \mathbf{P}_{\mu \nu}\cdot \mathbf{P}^{\mu \nu}\cr
&+&\Lambda_{\omega}\left(g^{2}_{\omega} \omega_{\mu}\omega^{\mu}\right)\left(g^{2}_{\rho} \boldsymbol{\rho}_{\mu}\cdot\boldsymbol{\rho}^{\mu}\right)  \cr
&+&\frac{1}{2}\left(\partial_{\mu}\sigma^* \partial^{\mu}\sigma^*
-m^{2}_{\sigma^*}{\sigma^*}^{2}\right) \cr
&+&\frac{1}{2} m^{2}_{\phi}\phi_{\mu}\phi^{\mu}
-\frac{1}{4} \Phi_{\mu \nu}\Phi^{\mu \nu}
\label{lagran}
\eea
where $D^{\mu}_{B}=i\partial^{\mu}-g_{\omega  B} \omega^{\mu}-g_{\phi
  B} \phi^{\mu}-g_{\rho B}\boldsymbol{\tau}_{B}\cdot\boldsymbol{\rho}^{\mu}$
and $m^{*}_{B}=m_{B}-g_{\sigma B}\sigma-g_{\sigma^* B}\sigma^*$ is
the effective mass of baryon $B$. $\Psi_{B}$ and $\psi_{l}$ are the baryon and
lepton Dirac fields, 
respectively, and $g_{iB}$ is the  coupling constant of  meson $i$ with baryon $B$.
The mass of baryon $B$ and  lepton $l$ are denoted by $m_{B}$ and
$m_{l}$, respectively. The constants $k$, $\lambda$, $\Lambda_\omega$
are the couplings associated with the non-linear interaction terms, and $\boldsymbol{\tau}_{B}$ is the isospin operator. The mesonic field tensors are given by their usual expressions: 
$\Omega_{\mu \nu}=\partial_{\mu}\omega_{\nu}-\partial_{\nu}\omega_{\mu}$,
$\boldsymbol{P}_{\mu \nu}=\partial_{\mu}\boldsymbol{\rho}_{\nu}-
\partial_{\nu}\boldsymbol{\rho}_{\mu}-g_\rho\left(\boldsymbol{\rho}_\mu\times \boldsymbol{\rho}_\nu\right)$, and $\Phi_{\mu \nu}=\partial_{\mu}\phi_{\nu}-\partial_{\nu}\phi_{\mu}$.
The couplings $g_{i,B}$ are constant for for the models TM1,
TM2$\omega\rho$, NL3 and NL3$\omega\rho$ whereas are density dependent in
DDME2. We will explain latter in section \ref{sect:calib} how all these couplings are fixed.
Here we simply indicate that the coupling constants of the nucleons with the $\sigma^*$ and
$\phi$ mesons are set to zero.

\section{Brief overview of hypernuclear physics}
\label{sect:Yphi}
Whereas the NN interaction is fairly well known due to the large number of existing scattering data, the YN and YY ones are still poorly constrained. Experimental difficulties due to the short lifetime of hyperons and the low intensity beam fluxes have limited the number of $\Lambda$N and $\Sigma$N scattering events to several hundreds \cite{engelmann66,alexander68,sechi68,kadyk71,eisele71}, and that of $\Xi$N events to very few. In the case of the YY interaction the situation is even worse because no scattering data exists at all. This limited amount of data is not enough to fully constrain these interactions. 

In the absence of scattering data, alternative information on the YN and YY interactions can be obtained from the study of hypernuclei, bound systems composed of nucleons and one or more hyperons. Hypernuclei were discovered in 1952 with the observation of a hyperfragment in a ballon-flown emulsion stack by Danysz and Pniewski \cite{dapnie52}. Since then more than 40 single $\Lambda$-hypernuclei, and few double$-\Lambda$ \cite{dl0,dl1,dl1b,dl2,dl3,dl4,dl5,nagara} and single$-\Xi$ \cite{khaustov00,nakazawa15} ones have been identified thanks to the use of high-energy accelerators and modern electronic counters. On the contrary, it has not been possible to prove without any ambiguity the existence of $\Sigma$-hypernuclei (see {\it e.g.,} Refs.\  \cite{bertini80,bertini84,bertini85,piekarz82,yamazaki85,tang88,bart99,hayano89,nagae98}) which suggests that the $\Sigma$-nucleon interaction is most probably repulsive \cite{dgm89,batty1,batty2,batty3,mares95,dabrowski99,noumi02,saha04,harada05,harada06}. 

Single $\Lambda$-hypernuclei can be produced by several mechanisms such as: $(K^-,\pi^-)$ {\it strangeness exchange reactions}, where a neutron hit by a $K^-$ is changed into a $\Lambda$ emitting a $\pi^-$. The analysis of these reactions showed many of the hypernuclear characteristics such as, for instance,  the small spin-orbit strength of the YN interaction, or the fact that the $\Lambda$ essentially retains its identity inside the nucleus. The use of $\pi^+$ beams permitted to perform $(\pi^+,K^+)$ {\it associated production reactions}, where an $s\bar s$ pair is created from the vacuum, and a $K^+$ and a $\Lambda$ are produced in the final state. The {\it electro-production} of hypernuclei by means of the reaction $(e,e'K^+)$ provides a high precision tool for the study of hypernuclear spectroscopy \cite{hugenford94} due to the excellent spatial and energy resolution of the electron beams. Recently, the HypHI collaboration at FAIR/GSI has proposed a new way to produce hypernuclei by using stable and unstable heavy ion beams \cite{hypHI}. The $\Lambda$ and the $^3_{\Lambda}$H and $^4_{\Lambda}$H hypernuclei have been observed in a first experiment performed using a $^6$Li beam on a $^{12}$C target at 2 A GeV \cite{rappold13}.

Hypernuclei can be produced in excited states if a nucleon in a p or higher shell is replaced by a hyperon. The energy of these excited states can be released either by emitting nucleons, or, sometimes, when the hyperon moves to lower energy states, by the emission of $\gamma$-rays. Measurements of $\gamma$-ray transitions in $\Lambda$-hypernuclei has allowed to analyse excited levels with an excellent energy resolution. Systematic studies of single $\Lambda$-hypernuclei indicate that the $\Lambda$N interaction is clearly attractive \cite{hashimoto06}.

$\Sigma$-hypernuclei can also be produced by the mechanisms just described. However, as said before, there is not yet an unambiguous experimental confirmation of their existence.

 To produce
double-$\Lambda$ hypernuclei, first it is necessary to create a $\Xi^-$ through reactions like
\begin{equation}
K^-\,\,+\,\,p\,\,\rightarrow\,\,\Xi^-\,\,+\,\,K^+ \ ,
\label{eq:xi1}
\end{equation}
or
\begin{equation}
p\,\,+\,\,\bar p\,\,\rightarrow\,\,\Xi^-\,\,+\,\,\bar \Xi^+ \ .
\label{eq:xi2}
\end{equation}
Then, the $\Xi^-$ should be captured in an atomic orbit and interact with the nuclear core producing two $\Lambda$ hyperons by means of the process
\begin{equation}
\Xi^-\,\,+\,\,p\,\,\rightarrow\,\,\Lambda\,\,+\,\,\Lambda\,\,+\,\,28.5\,\,\mbox{MeV}  \ ,
\end{equation}
providing about $30$ MeV of energy that is equally shared between the two $\Lambda$'s in most cases, leading to the escape of one or both hyperons from the nucleus. $\Xi$-hypernuclei can be produced by means of the reactions (\ref{eq:xi1}) and (\ref{eq:xi2}) and, as said above, very few of them have been identified. The analysis of the experimental data from production reactions such as $^{12}$C$(K^-,K^+)^{12}_{\Xi^-}$Be \cite{kau} indicates an attractive $\Xi$-nucleus interaction of the order of about $\sim -14$ MeV. Here
we should mention the very recent observation of a deeply bound state of the $\Xi^-$-$^{14}$N system with a binding energy of $4.38\pm 0.25$ MeV  by
Nakazawa {\it et al.} \cite{nakazawa15}. This event provides the first clear evidence of a deeply bound state of
this system by an attractive $\Xi$N interaction. Future $\Xi$-hypernuclei experiments are being planned at J-PARC.

Double-strange hypernuclei are nowadays the best systems to investigate the properties of the 
baryon-baryon interaction in the strangeness $S=-2$ sector . The 
$\Lambda\Lambda$ bond energy $\Delta B_{\Lambda\Lambda}$ in double-$\Lambda$ hypernuclei can be determined experimentally from the measurement of the binding energies of double- and single-$\Lambda$ hypernuclei as
\begin{equation}
\Delta B_{\Lambda\Lambda}=B_{\Lambda\Lambda}(^A_{\Lambda\Lambda}Z)-2B_{\Lambda}(^{A-1}_{\Lambda}Z) \ .
\end{equation}

Emulsion experiments \cite{dl1,dl2,dl3,dl4} have reported the formation of a few double-$\Lambda$ hypernuclei: $^6_{\Lambda\Lambda}$He, $^{10}_{\Lambda\Lambda}$Be and $^{13}_{\Lambda\Lambda}$B. From the subsequent analysis of these emulsion experiments a quite large $\Lambda\Lambda$ bound energy of around 4 to 5 MeV was deduced, contrary to expectation from SU(3) (Stoks and Rijken 1999 in Ref.\ \cite{nijmegen}). We should also note that the identification of some of these double-$\Lambda$ hypernuclei was ambiguous. Therefore, careful attention should be paid when using the data from this old analysis to put any kind of constraint on the $\Lambda\Lambda$ interaction. However, a new $_{\Lambda\Lambda}^6$He candidate having a $\Lambda\Lambda$ bond energy
\begin{equation}
\Delta B_{\Lambda \Lambda}=1.01\pm 0.2^{+0.18}_{-0.11} \,\,\rm{MeV}
\label{eq:tak}
\end{equation}
was unambiguously observed in 2001 at KEK \cite{nagara}. This value has then been recently revised due to a change in the value of the $\Xi^-$ mass \cite{Ahn13}: 
\begin{equation}
\Delta B_{\Lambda \Lambda}=0.67\pm 0.17\,\,\rm{MeV}.
\label{eq:ahn}
\end{equation}
In this work we will use these two values of $\Delta B_{\Lambda\Lambda}$ to constrain the coupling of the $\Lambda$ hyperon with the $\sigma^*-$meson.

\section{Calibration of the $\Lambda-$meson coupling constants}
\label{sect:calib}

Since the $\Lambda$ is an isospin-singlet it does not couple with the $\rho-$meson. Therefore, only the coupling constants with the $\sigma-, \omega-, \sigma^*-$ and $\phi-$mesons should be fixed. The usual
procedure to fix these couplings consists in using the SU(6) symmetry to determine the couplings of the $\Lambda$ with the vector mesons in terms of those of the nucleons
\bea
 R_{\omega\Lambda}=g_{\omega\Lambda}/g_{\omega N}=2/3
\eea
\bea
 R_{\phi\Lambda}=g_{\phi\Lambda}/g_{\omega N}=- \frac{\sqrt{2}}{3},
\eea
and the $\Lambda$-scalar mesons ones by using data derived indirectly from hypernuclei. In particular, these couplings are obtained by imposing the value of the  $\Lambda$ potential in symmetric nuclear matter, $U_{\Lambda}^{N}$, and the value of the $\Lambda$-potential in $\Lambda$-matter, $U_{\Lambda}^{\Lambda}$,  at saturation, defined respectively as: 
\bea
 U_{\Lambda}^{N}(n_0)&=
	- \left(g_{\sigma \Lambda} + g'_{\sigma \Lambda} \rho_s\right)\sigma_0 
	+ \left(g_{\omega \Lambda}+  g'_{\omega \Lambda} n_0\right)\omega_0,
	\label{eq:potential}
\eea
and
\bea
	U_{\Lambda}^{\Lambda}(n_0)=&
	-\left( g_{\sigma \Lambda} + g'_{\sigma \Lambda} \rho_s\right)\sigma_0
	-\left( g_{{\sigma^{\ast}} \Lambda} + g'_{\sigma^* \Lambda} \rho_s\right) \sigma^*_0\nonumber\\
	&+ \left( g_{\omega \Lambda} +  g'_{\omega \Lambda} n_0\right)\omega_0
	+ \left(g_{\phi \Lambda} + g'_{\phi \Lambda} n_0\right)\phi_0 \ . 
	\label{eq:potential-ss}
\eea
with $\sigma_0$, $\omega_0$, $\sigma^*_0$, and $\phi_0$  the mean-field values of the $\sigma$, $\omega$, $\sigma^*$, and $\phi$ meson fields, respectively, and $\rho_s$ the scalar density. The quantities $g'_{i\Lambda}$ 
are the derivatives with respect to the density of the couplings $g_{i \Lambda}$ 
and are only different from zero for models with density dependent couplings (see the discussion below). All quantities are calculated at the saturation density $n_0$. 
Values of $U_{\Lambda}^{N}(n_0) \simeq -30$ MeV and $U_{\Lambda}^{\Lambda}(n_0) =-5$ MeV are usually employed in the literature to determine these couplings. The first value results from the extrapolation at $A^{-2/3}=0$ of the experimental binding energy of single-$\Lambda$ hypernuclei, $A$ being the mass number of the hypernucleus. The second one is usually obtained from the identification $U_{\Lambda}^{\Lambda}(n_0)=-\Delta B_{\Lambda\Lambda}$ and the use of a value of 5 MeV for the binding energy of two $\Lambda$'s. However, as pointed before, one has to be very careful when using this old experimental
data. 

In this work, however, we follow a different procedure. The couplings of the $\Lambda$ with the various mesons are calibrated by fitting the experimental binding energy of $\Lambda$-hypernuclei following the approach of Refs.\ \cite{Sugahara94,Shen06}. Before we give more specific details on the calibration procedure, we should note that we have considered two different approaches to fix the hyperon-meson couplings in the case of the model DDME2. First, we use the experimental constraints and symmetry arguments to fix the magnitude of the couplings at saturation density, as done for the other models with constant couplings. Then we consider: (i) that the hyperon-meson couplings do not depend on the density, this approach is designated simply as DDME2; and (ii) we assume for the hyperon couplings the same density dependence of the nucleonic couplings, this model will be referred as DDME2D. For explicit density dependence of the couplings the interested reader is referred to Ref.\ \cite{DDME2}.

Hypernuclei binding energies are obtained by solving the
 Dirac equations for the nucleons and the $\Lambda$ obtained from the
  Lagrangian density (\ref{lagran}) 
 using the method described in Refs.\
  \cite{Avancini07,GRT}. In this approach the hypernucleus wave function
  is a Slater determinant  and only the lowest single-particle positive energy states
 are occupied. 
  We use the relativistic mean field approximation where the meson field operators are replaced by their 
  expectation values 
  and negative energy states are neglected (no-sea approximation). 
  The numerical algorithm consists in the expansion of the Dirac spinors and mesonic fields 
  in terms of the harmonic oscillator basis. Therefore, the Dirac and Klein-Gordon equations are 
  transformed into matrix equations that are solved in a self-consistent way until convergence is achieved. 
  For an accurate description
  of light hypernuclei the center-of-mass correction, instead of the simple correction, 
\begin{equation}
E_{COM}=\frac{3}{4}41A^{-1/3}\,\,\,\, \mbox{[MeV]}\ , 
\end{equation}
commonly used in the literature \cite{GRT},
is calculated through the expression,
\begin{equation}
 E_{COM}= \frac{\left\langle P^2 \right\rangle }{2M}\ ,
\end{equation}
 where   $\left\langle P^2 \right\rangle$ is the expectation value of the squared total momentum and $M$ 
is the hypernucleus total mass. The former expectation value is calculated from the actual many-body state of the hypernucleus.

  As in Ref.\ \cite{Shen06} we include the  tensor term 
\begin{equation}
{\cal L}_{T\Lambda}=
\bar\psi_\Lambda \frac{f_{\omega\Lambda}}{2M_\Lambda}\sigma^{\mu\nu}\partial_\nu\omega_\mu\ \psi_\Lambda \ ,
\end{equation}
which is important to get a weak $\Lambda$-nuclear spin-orbit
interaction \cite{noble80,jennings91}. The spin-orbit potential for single-$\Lambda$ hypernuclei 
is the result of two opposite contributions which partially cancel out, one is the usual associated to the difference
between the derivative of the scalar ($\sigma$)   and vector ($\omega$) central potentials and  
the other due to the tensor term. 

Although no experimental data are available for the spin-orbit splitting of $\Lambda$ hypernuclei, 
taking into account the tensor term, within the quark model ($f_{\omega\Lambda}=-g_{\omega \Lambda}$),
causes an improvement on the quality of the overall calibration of the coupling constants.

\subsection{Single $\Lambda$-hypernuclei}
\begin{figure}
	\includegraphics[width=\columnwidth]{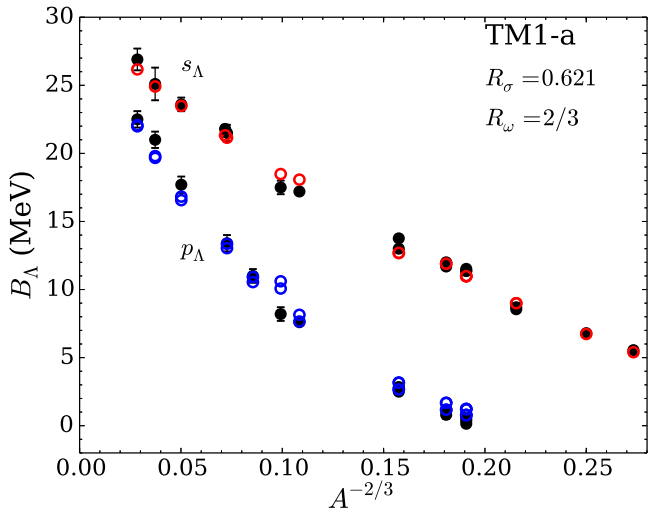}

    \caption{ For the TM1-a model, experimental values of the binding energies $B_\Lambda$ in the s- and p-shells of single $\Lambda$-hypernuclei (black circles) and modelled values (red and blue open circles, respectively) obtained after adjusting $R_{\sigma\Lambda}$ in order to minimize the quantity $\chi^2$ defined in Eq.\,(\ref{eq:chi}). For $\Lambda$ in p-shells, p$_{1/2}$ and p$_{3/2}$ states are plotted.}
    \label{fig:Calib}
\end{figure}

For a given value of $R_{\omega\Lambda}$, the ratio $R_{\sigma\Lambda}=g_{\sigma\Lambda}/g_{\sigma N}$ is calibrated to reproduce the binding energies $B_\Lambda$ of hypernuclei in the s- and p-shells (see Figure \ref{fig:Calib}). The experimental data used in the calibration is taken from Table IV of Ref.\ \cite{Gal16}. The best value of $R_{\sigma\Lambda}$ is determined by minimizing the function:
\be
\chi^2=\frac{1}{N}\sum_{i=1}^N \left(\frac{B_{\Lambda_ i}^{\rm exp}-B_{\Lambda_ i}^{\rm the}}{B_{\Lambda_ i}^{\rm exp}}\right)^2
\label{eq:chi}
\ee
where $B_{\Lambda i}^{\rm exp}$ and $B_{\Lambda i}^{\rm the}$ are, respectively, the values of the binding energy  of a given single $\Lambda$-hypernuclei $i$ obtained experimentally and from the modelling, and $N$ is the total number of $\Lambda$-hypernuclei for which experimental data is available. Equal, or very close, values of $R_{\sigma\Lambda}$ are obtained if only heavy hypernuclei with $Z>20$ are considered or if the denominator in Eq.~(\ref{eq:chi}) is replaced by the error bar on the experimental measurements of the binding energies. Similarly the calibration is hardly affected if only s-shell binding energies are taken into account or if both shells are included. In Table~\ref{tab:single} we indicate, for two different values of $R_{\omega\Lambda}$, the calibrated values of $R_{\sigma\Lambda}$ as well as the associated value of the $\Lambda$-potential in symmetric baryonic matter at saturation $U_\Lambda^N(n_0)$ obtained from Eq.~(\ref{eq:potential}), for all the models considered. In this Table and  in the following for each RMF parametrization we consider two values of the ratio $R_{\omega\Lambda}$, one $R_{\omega\Lambda}=2/3$ corresponding to SU(6) symmetry case labelled `a' and a second with $R_{\omega\Lambda}=1$ labelled `b', where the symmetry is broken.  We note that the values of the couplings and that of $U_\Lambda^N(n_0)$ in Table~\ref{tab:single} are remarkably similar: $R_{\sigma\Lambda}\simeq 0.62$ for the a- models 
and $R_{\sigma\Lambda} \simeq 0.89$ for the b- models, and $U_\Lambda^N(n_0)\simeq -(30$ -- $32)$
MeV for all the models except three of them. In Figure~\ref{fig:Calib} the experimental values and the theoretical ones obtained after calibration are plotted for the TM1 model with $R_{\omega\Lambda}=2/3$.

\begin{table}
\begin{ruledtabular}
\center\begin{tabular}{cccc}
Model & $R_{\omega\Lambda}$ & $R_{\sigma\Lambda}$  & $U_\Lambda^N(n_0)$\\ 
\hline
TM1-a & 2/3 & 0.621 &  -30 \\ 
TM1-b & 1   & 0.892 &  -31 \\ 
&&&\\
TM2$\omega\rho$-a & 2/3 & 0.624 &  -31 \\ 
TM2$\omega\rho$-b & 1   & 0.905 &  -36 \\ 
&&&\\
NL3-a & 2/3 & 0.622 & -31 \\ 
NL3-b & 1   & 0.894 & -32 \\ 
&&&\\
NL3$\omega\rho$-a & 2/3 & 0.622 & -31 \\ 
NL3$\omega\rho$-b & 1   & 0.894 & -32 \\ 
&&&\\
DDME2-a & 2/3 & 0.615 & -32 \\
DDME2-b & 1   & 0.891 & -35 \\
&&&\\
DDME2D-a & 2/3 & 0.621 & -32 \\
DDME2D-b & 1   & 0.896 & -35 \\
 \end{tabular} 
\caption{Calibration to single $\Lambda$-hypernuclei: for given $R_{\omega\Lambda}$, values of $R_{\sigma\Lambda}$ calibrated to reproduce the binding energies $B_\Lambda$ of hypernuclei in the $s$ and $p$ shells. The last column contains the value of the $\Lambda$-potential in symmetric baryonic matter at saturation in MeV, for reference. }
\label{tab:single}
\end{ruledtabular}
\end{table}

\subsection{Double $\Lambda$-hypernuclei}


\begin{figure*}
	\includegraphics[width=\columnwidth]{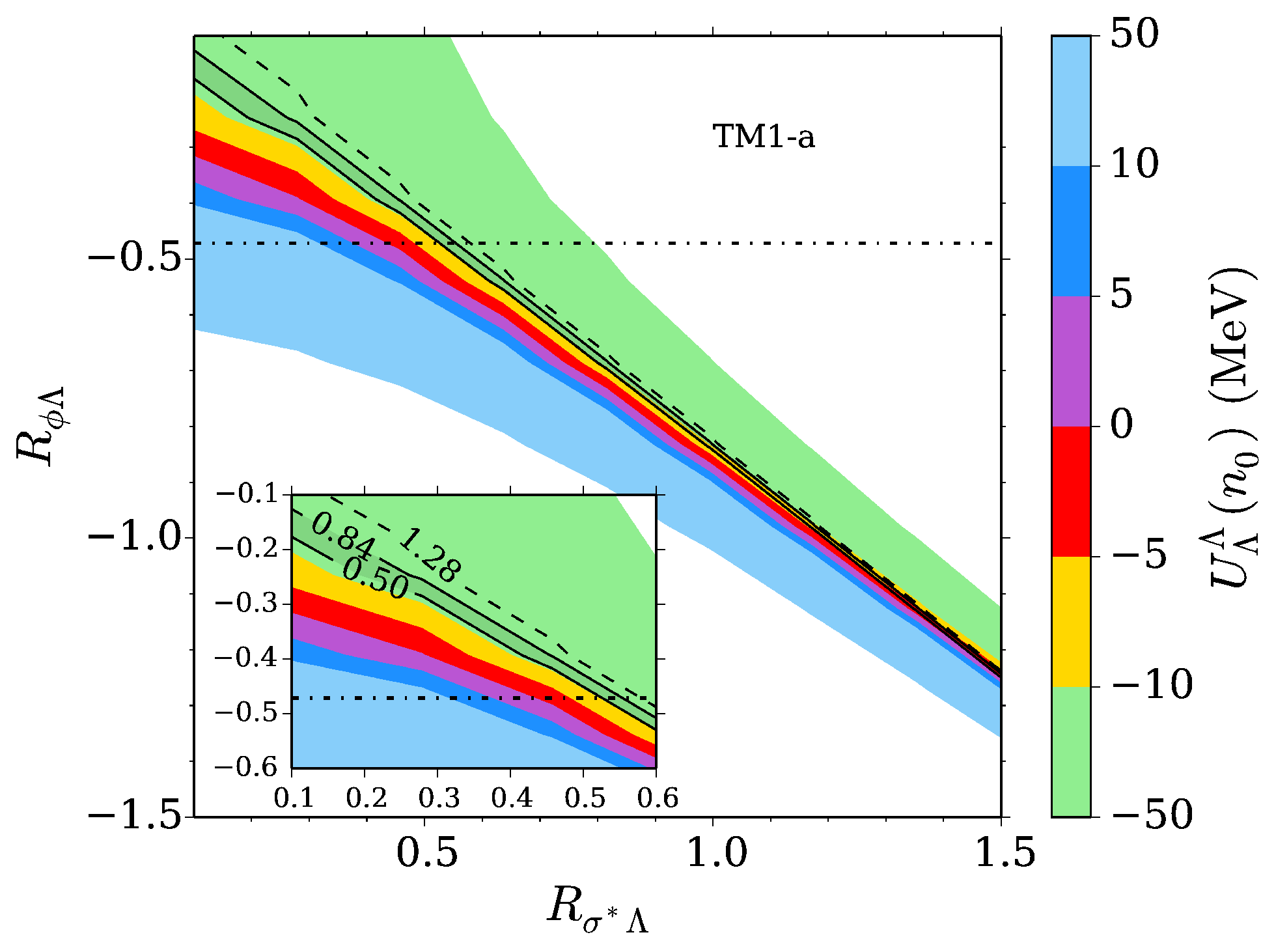}
	\includegraphics[width=\columnwidth]{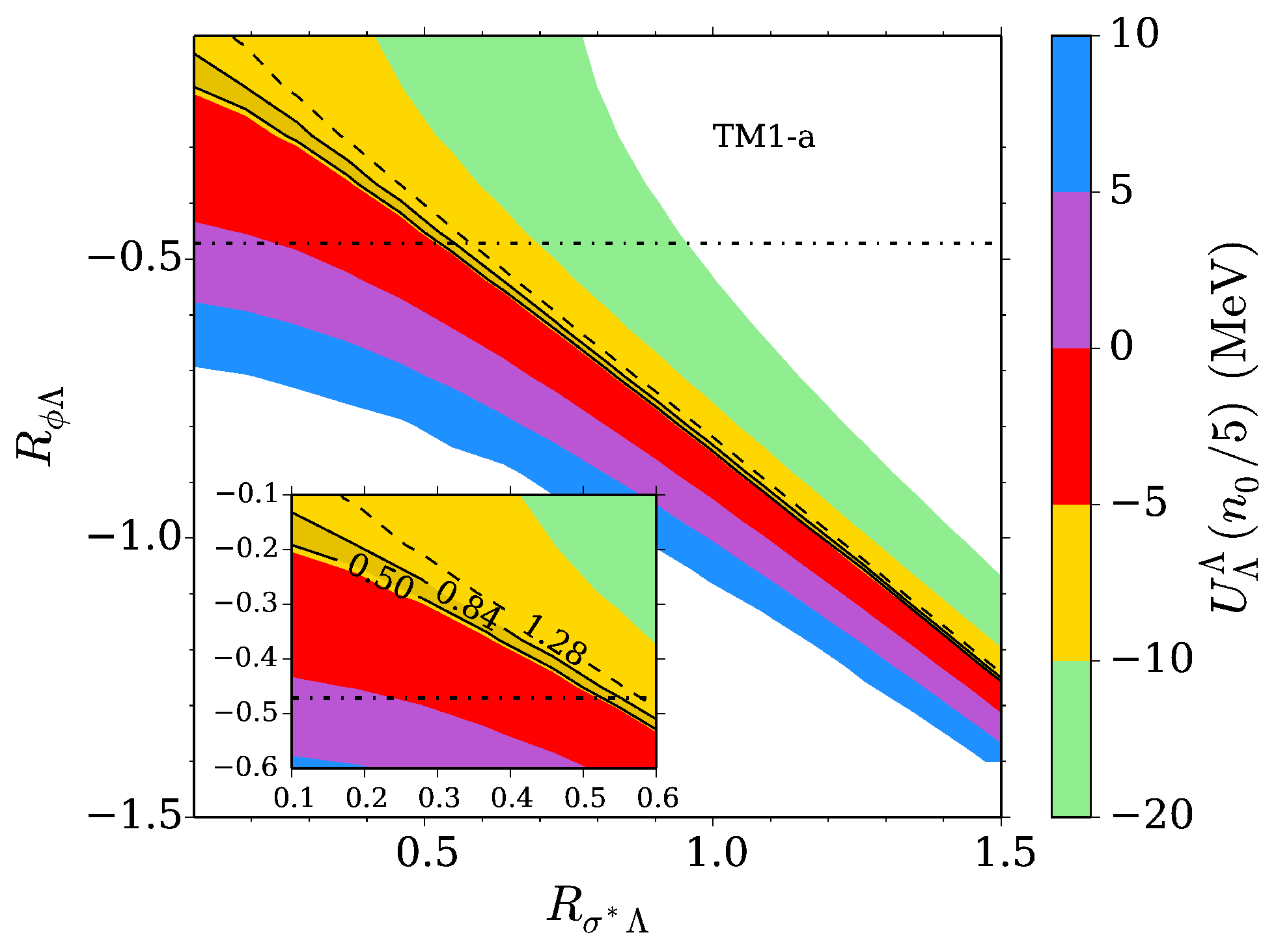}
	\includegraphics[width=\columnwidth]{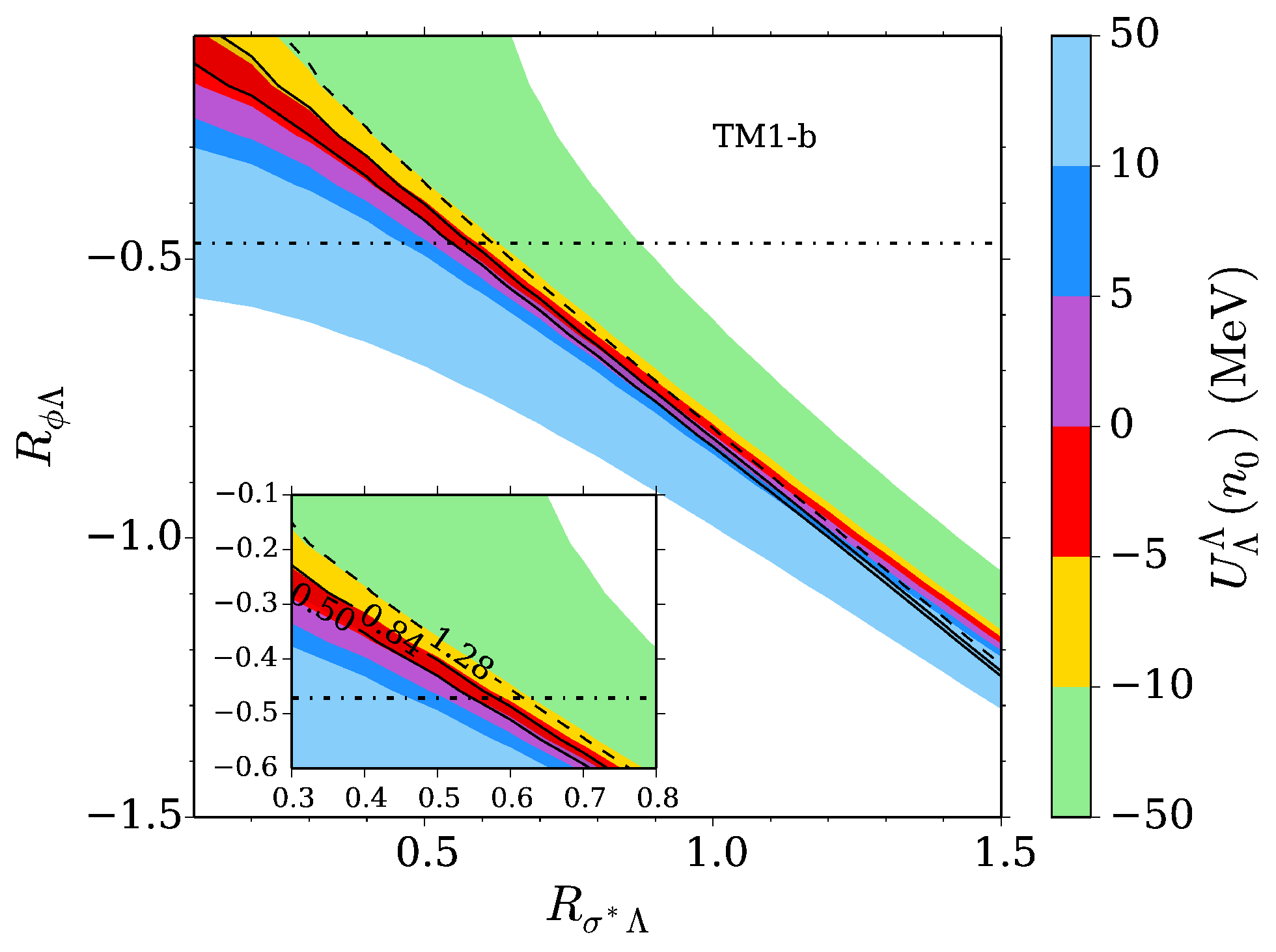}
	\includegraphics[width=\columnwidth]{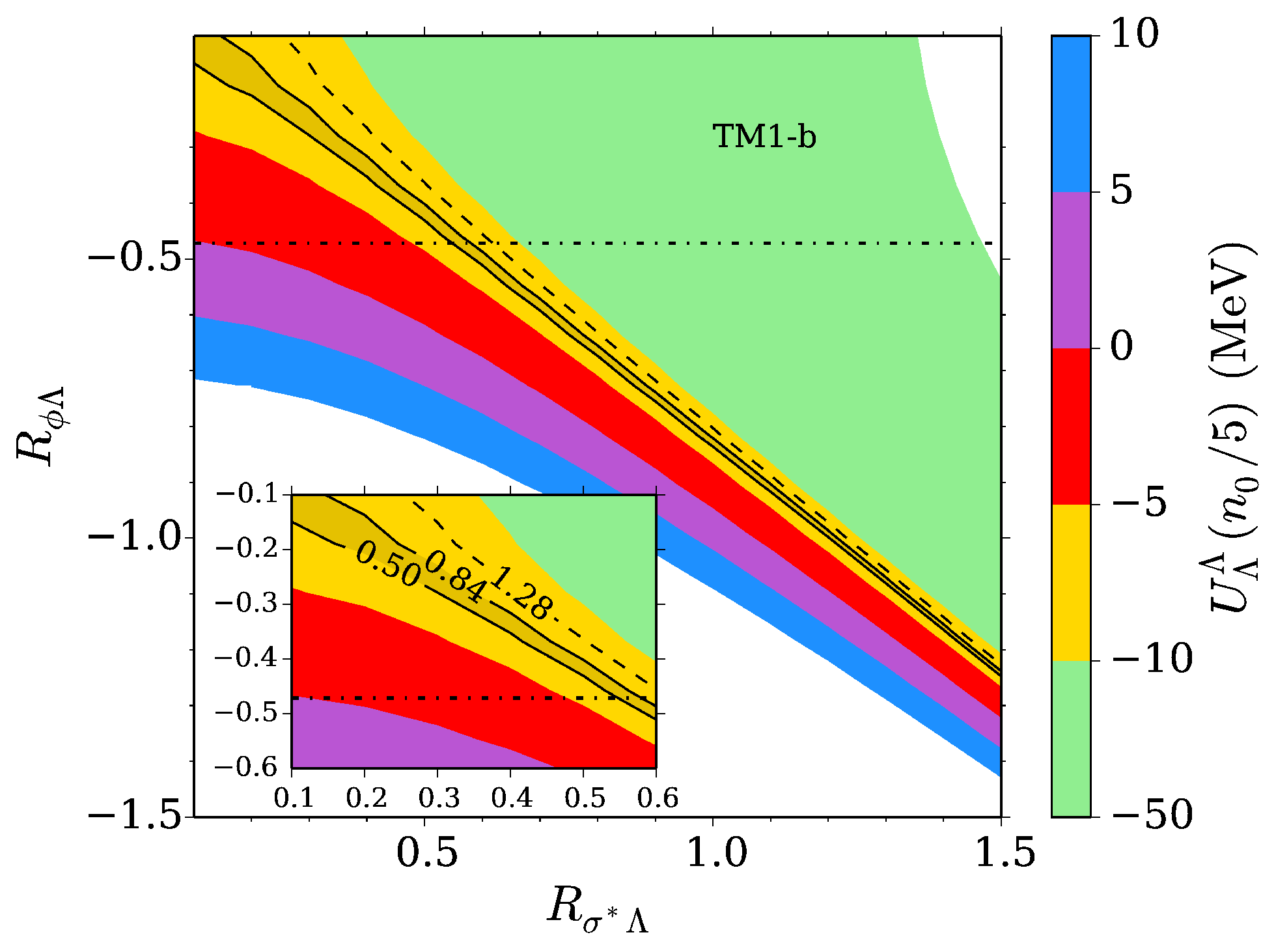}
    \caption{TM1-a (top) and -b (bottom) models. The solid and dashed black lines correspond to values of $(R_{\sigma^{\ast}\Lambda},R_{\phi\Lambda})$ consistent with the experimental values of the bound energy of $_{\Lambda \Lambda}^6$He in Eqs.\,(\ref{eq:tak}-\ref{eq:ahn}). Colour contours: values of $U_\Lambda^\Lambda$ at saturation density $n_0$ (left) and at $n_0/5$ (right) obtained from Eq.\,(\ref{eq:potential-ss}).}
    \label{fig:TM1:BLL_ULL}
\end{figure*}

\begin{table*}
\begin{ruledtabular}
\center\begin{tabular}{cc|ccc|ccc}
Model &    & \multicolumn{3}{c|}{$\Delta B_{\Lambda \Lambda}=0.50$} & \multicolumn{3}{c}{$\Delta B_{\Lambda \Lambda}=0.84$}\\
&  $R_{\phi\Lambda}$ &$R_{\sigma^\ast\Lambda}$ &  $U_\Lambda^\Lambda(n_0)$ & $U_\Lambda^\Lambda(n_0/5)$ &$R_{\sigma^{\ast}\Lambda}$& $U_\Lambda^\Lambda(n_0)$& $U_\Lambda^\Lambda(n_0/5)$ \\ 
\hline
TM1-a & $-\sqrt{2}/3$ & 0.533 & -11.2 & -5.3 & 0.557 & -14.2 & -5.9\\ 
      & $-\sqrt{2}/2$ & 0.833 & -10.0 &  -5.4 & 0.849 & -13.0 & -6.0 \\ 
TM1-b & $-\sqrt{2}/3$ & 0.549 & -0.2  & -6.8 & 0.580 & -4.1 & -7.6 \\
      & $-\sqrt{2}/2$ & 0.843 &  2.7  & -6.8 & 0.864 & -1.2 & -7.7 \\

&&&\\
TM2$\omega\rho$-a & $-\sqrt{2}/3$ & 0.547 & -13.5 & -5.9 & 0.567 & -16.0 & -6.4 \\ 
      & $-\sqrt{2}/2$ & 0.850 & -14.2 & -6.4 & 0.863 & -16.6 & -6.9\\ 
TM2$\omega\rho$-b & $-\sqrt{2}/3$ & 0.563 & -9.4 & -9.4 & 0.589 & -12.7 & -10.1\\
      & $-\sqrt{2}/2$ & 0.859 & -8.2 & -9.8 & 0.877 & -11.4 & -10.5\\

&&&\\
NL3-a & $-\sqrt{2}/3$ & 0.534 & -9.9 & -5.6 & 0.559 & -13.2 & -6.3 \\ 
      & $-\sqrt{2}/2$ & 0.835 & -8.4 & -5.7 & 0.851 & -11.6 & -6.4\\ 
NL3-b & $-\sqrt{2}/3$ & 0.552 & 5.2 & -7.0 & 0.586 & 0.8 & -8.0 \\ 
      & $-\sqrt{2}/2$ & 0.846 & 9.0 & -7.1 & 0.868 & 4.8 & -8.0\\ 
&&&\\
NL3$\omega\rho$-a & $-\sqrt{2}/3$ & 0.534 & -9.4 & -5.5 & 0.560 & -12.8 & -6.2 \\ 
                  & $-\sqrt{2}/2$ & 0.835 & -7.9 & -5.6 & 0.851 & -11.2 & -6.3 \\ 
NL3$\omega\rho$-b & $-\sqrt{2}/3$ & 0.552 & 5.2 & -7.0 & 0.586 & 0.8 & -8.0 \\ 
                  & $-\sqrt{2}/2$ & 0.846 & 9.0 & -7.1 & 0.868 & 4.8 & -8.0 \\ 
&&&\\
DDME2-a & $-\sqrt{2}/3$ & 0.538 & -8.4&-2.7 & 0.561 &-11.6&-3.4\\ 
        & $-\sqrt{2}/2$ & 0.828 &-6.2&-2.7 & 0.843 &-9.4&-3.4 \\ 
DDME2-b & $-\sqrt{2}/3$ & 0.563 & 1.6&-2.8& 0.592 & -2.5&-3.7\\ 
        & $-\sqrt{2}/2$ & 0.844 & 6.6&-2.8& 0.864 & 2.6&-3.7\\ 
&&&\\
DDME2D-a & $-\sqrt{2}/3$ & 0.535 & -11.9&-4.1& 0.555 & -11.7&-4.0\\ 
         & $-\sqrt{2}/2$ & 0.826 & -10.6&-4.0& 0.840 & -10.6&-4.0\\ 
DDME2D-b & $-\sqrt{2}/3$ & 0.564 &  -6.7&-4.3& 0.588 &-6.6&-4.3\\ 
         & $-\sqrt{2}/2$ & 0.846 &-3.4&-4.3  & 0.862 &-3.4&-4.3\\ 
         
\end{tabular} 
\caption{Calibration to double $\Lambda$-hypernuclei for all models a and b. For given $R_{\phi\Lambda}$, $R_{\sigma^{\ast}\Lambda}$ are calibrated to reproduce the upper and lower values of bound energy of $_{\Lambda \Lambda}^6$He. For reference the $\Lambda$-potential in pure $\Lambda$-matter at saturation and at $n_0/5$ are also given. All energies are given in MeV.}
\label{tab:double}
\end{ruledtabular}
\end{table*}


The value of the coupling constants of the $\Lambda$ to the hidden-strangeness mesons $\sigma^*$ and $\phi$ is calibrated using the measured $\Lambda \Lambda$ bond energy of $_{\Lambda \Lambda}^6$He.
Figure~\ref{fig:TM1:BLL_ULL} shows for the TM1-a and -b models (with $R_{\sigma\Lambda}$ values adjusted to single $\Lambda$-hypernuclei) lines of constant $\Delta B_{\Lambda \Lambda}$ 
consistent with the experimental values of the bound energy of $_{\Lambda \Lambda}^6$He, {\it i.e.,} within 
the error bars defined in Eqs.\ (\ref{eq:tak})  and (\ref{eq:ahn}). 
In particular, the continuous lines correspond to the limits of Eq. (7),
and the dashed line to the upper limit of Eq.\ (\ref{eq:tak}), the lower limit being
inside the interval defined by Eq. (\ref{eq:ahn}). The color contours for the figures on the left side indicate the value of the $\Lambda$-potential in $\Lambda$-matter at saturation, $U_{\Lambda}^{(\Lambda)}(n_0)$, obtained from Eq.\,(\ref{eq:potential-ss}). For completeness in the figures on the right side we also plot contours for $U_{\Lambda}^{(\Lambda)}$ at $n_0/5$ as this is the quantity that has been used to determine the couplings {\it e.g.,} in Ref. \cite{Oertel14}. The horizontal line corresponds to the SU(6) value $R_{\phi\Lambda}=-\sqrt{2}/3$.  We note that for most values of the ratios $R_{\sigma^{\ast}\Lambda}=g_{\sigma^{\ast}\Lambda}/g_{\sigma N}$ and $R_{\phi\Lambda}$, consistent with the experimental constraint from $_{\Lambda \Lambda}^6$He, the value of  $U_{\Lambda}^{(\Lambda)}(n_0)$ potential, however, greatly varies and is very different from the value of $\sim -5$ MeV generally  used in the literature to fix these couplings. In Table \ref{tab:double} we indicate the values of $R_{\sigma^{\ast}\Lambda}$, $U_{\Lambda}^{(\Lambda)}(n_0)$, and $U_{\Lambda}^{(\Lambda)}(n_0/5)$ for two choices of the ratio $R_{\phi\Lambda}$: the one corresponding the SU(6) symmetry, $R_{\phi\Lambda}=-\sqrt{2}/3\simeq-0.471$, and  another one $R_{\phi\Lambda}=-\sqrt{2}/2\simeq-0.707$ for which the symmetry is broken. The values of $R_{\sigma^{\ast}\Lambda}$ and $R_{\phi\Lambda}$ have been obtained after calibrating to the lower and upper values of the $_{\Lambda \Lambda}^6$He binding energy given in Eq.\ (\ref{eq:ahn}), $\Delta B_{\Lambda \Lambda}=0.50$ and
$\Delta B_{\Lambda \Lambda}=0.84$~MeV, respectively, for our set of models. On the one hand $U_{\Lambda}^{(\Lambda)}$ at saturation is shown to vary from $\sim -16$ to $-8$ MeV taking the SU(6) values for vector mesons couplings, and  $-17$ and $+9$ MeV if the vector meson couplings to
hyperons are imposed to be not larger than to nucleons. These ranges strongly differ from the generally employed one in the literature, {\it i.e.,} between $-1$ and $-5$ MeV, showing that the use of such values for $U_{\Lambda}^{(\Lambda)}$  is inconsistent with the hypernuclei data. On the other the values of $U_{\Lambda}^{(\Lambda)}$ evaluated at $n_0/5$ are restricted to a smaller range: $-5$ to $-11$ MeV approximately. On the whole, Tables \ref{tab:single} and \ref{tab:double} provide the complete set of values of the coupling constants for the $\Lambda$ calibrated to hypernuclear data for all our parametrizations.

\section{Hyperonic Neutron Stars}
\label{sect:NS}

We now explore how the calibration of the coupling constants for the $\Lambda$ hyperon to the binding energies of single and double $\Lambda$-hypernuclei affects the properties of neutron stars, in particular the maximum mass. To do so we calculate the EoS for neutron star matter. Following the work of two of the authors \cite{Fortin16} unified EoS are built.  For the outer crust we take the  EoS proposed in Ref.\ \cite {ruester06}, for the inner crust we perform a Thomas Fermi calculation and allow for
non-spherical clusters according to \cite{GP0,GP} and for the core we
consider the homogeneous matter EoS. 

It is well know that the consequence of the inclusion of hyperons is a softening of the EOS and thus a reduction of its maximum mass $M_{\rm max}$ compared to the purely nucleonic case. The more hyperonic species at the density corresponding to the central one of the NS with the maximum mass, the smaller the value of $M_{\rm max}$. Consequently, we consider two types of hyperonic models for the 
neutron star core: (i) a model in which in addition to the nucleons only the $\Lambda$ hyperon is present, and
(ii) a second one where we allow for the appearance of all the hyperon species from  the baryonic octet. The first model constitutes a ``minimal hyperonic model'' in the sense that only $\Lambda$, if they appear, are present at high density and, therefore, compared to models with a richer hyperonic composition it will predict largest maximum masses. Thus it defines the upper limit on the maximum mass of an hyperonic neutron star.

For the second model, in principle a procedure similar to the one presented in the previous section for $\Lambda$ could be used to determine the couplings for $\Sigma$ and $\Xi$ hyperons with the different mesons. However,  as mentioned in section \ref{sect:Yphi}, there is not yet an unambiguous experimental confirmation of the existence of the $\Sigma$-hypernuclei and very few $\Xi$-hypernuclei has been observed. Hence the couplings of the $\sigma-$meson to the $\Xi$ and $\Sigma$ hyperons cannot be calibrated using hypernuclear data.  Therefore, in this case, we fix the value of the single-particle potentials of the $\Sigma$ and $\Xi$ and use equations equivalent to Eq.\ (\ref{eq:potential}) to determine these couplings. In order to explore the dependence of the neutron star maximum mass on the choice of potentials we choose a repulsive potential for the $\Sigma$ hyperons: $U_\Sigma^N(n_0)=0, +30$ MeV and $U_\Xi^N(n_0)=-14$ MeV or $U_\Xi^N(2n_0/3)=-14$ MeV as suggested by the observations of $\Xi$-hypernuclei \cite{khaustov00,Gal16}. In addition, since double-$\Xi$ or double-$\Sigma$ hypernuclei
has not been observed, in this work we do not include the coupling of these hyperons with the $\phi-$ and $\sigma^{\ast}-$mesons. We adopt the SU(6) values for the couplings to vector-isoscalar mesons:
\bea
	&&g_{\omega\Xi}=\frac{1}{3} g_{\omega N} = \frac{1}{2} g_{\omega\Sigma} \ ,
		\label{eq:SU6-relation2}
\eea
\bea
         && g_{\phi\Xi} = 2 g_{\phi\Sigma} =- \frac{2\sqrt{2}}{3} g_{\omega N} 
	\label{eq:SU6-relation2b}
\eea

and assume 
\bea	
g_{\rho\Xi} = \frac{1}{2} g_{\rho\Sigma} = g_{\rho N}  
	\label{eq:SU6-relation3}
\eea
for the $\rho$-meson taking into account the isospin properties of the different baryons.

\begin{table}
\begin{ruledtabular}
\centering\begin{tabular}{cccccc}
$R_{\sigma\Lambda}$ &$R_{\sigma\Sigma}$&$R_{\sigma\Xi}$& $U_{\Lambda}$&$U_{\Sigma}$&$U_{\Xi}$\\
& & & (MeV)&MeV)&(MeV)\\
\hline
0.6164& 0.15& 0& -32.6&154.9&107.7\\
0.6164& 0.45 & 0.15& -32.6 &34.3 &47.5\\
0.6164& 0.76& 0.30& -32.6& -88.6&-14.0\\
\end{tabular}
\caption{Hyperon-$\sigma$ coupling ratios and hyperonic potentials in symmetric nuclear matter at saturation, using the conditions defined in \cite{Sedrakian13,Sedrakian14}.}
\label{sedrakian1}
\end{ruledtabular}
\end{table}

Here we should mention that the authors of Ref.\ \cite{Sedrakian13,Sedrakian14} have considered the SU(3) flavour symmetric model to fix the couplings of the hyperons to the three mesons, $\sigma$, $\omega$, and $\rho$, and have obtained for the last two the same couplings we define in Eqs.\ (\ref{eq:SU6-relation2})-(\ref{eq:SU6-relation3}). For the $\sigma$-hyperon couplings they arrive at the equality
\begin{equation}
2\left(g_{N\sigma}+ g_{\Xi\sigma}\right)=3g_{\Lambda\sigma}+ g_{\Sigma\sigma}.
\label{sedrakian}
\end{equation}
which they complement with two extra conditions imposing that the hyperon couplings are positive and smaller than the nucleon ones \cite{Sedrakian13}. From $\Lambda$-hypernuclei, the ratio $R_{\sigma\Lambda}$ was fixed  to 0.616 \cite{Sedrakian14} for the DDME2 model.  This value of $R_{\sigma\Lambda}$ together with the relation (\ref{sedrakian}) and the condition  $0\le g_{\Xi\sigma}\le g_{N\sigma}$ results in the following range of values for $R_{\sigma\Sigma}$: $0.15\le R_{\sigma\Sigma}\le 0.45$. Using these values for the meson-hyperon coupling ratios one can determine the hyperonic potentials in symmetric nuclear matter, taking  the  hyperon coupling parameters  constant. Although not indicated, this seems to have been the choice in \cite{Sedrakian13,Sedrakian14} since Eq.\ (33) in \cite{Sedrakian13} applies to constant couplings. The results are shown in Table \ref{sedrakian1}. The first two lines of this table have been obtained taking the lower and upper values for the ratio $R_{\sigma\Sigma}$. While the value  of $U_\Lambda(n_0)$ is within the expected range since $R_{\sigma\Lambda}$
was fitted to the $\Lambda$-hypernuclei properties, the $\Xi$ potential comes very repulsive contrary to the experimental results which seem to indicate that $\sim -14$ MeV would be a reasonable value \cite{khaustov00,Gal16}. Keeping now the $R_{\sigma\Lambda}$ ratio and choosing $R_{\sigma\Xi}$ such that $U_\Xi(n_0)=-14$ MeV,  
Eq.\ (\ref{sedrakian}) can be used to determine $R_{\sigma\Sigma}$. The results are shown in the last  line of Table \ref{sedrakian1}. One immediately sees that the $\Sigma$ potential comes out very attractive when, in fact, it is expected to be repulsive. It appears, therefore, that the constraints resulting from the SU(3) flavour symmetric model  for the hyperon-scalar-meson coupling constants are not compatible with a simultaneous attractive $\Xi$ potential and a repulsive $\Sigma$ potential, and thus are in contradiction with what experiments seem to indicate, as discussed in section III. 

\begin{figure*}
	\includegraphics[width=\columnwidth]{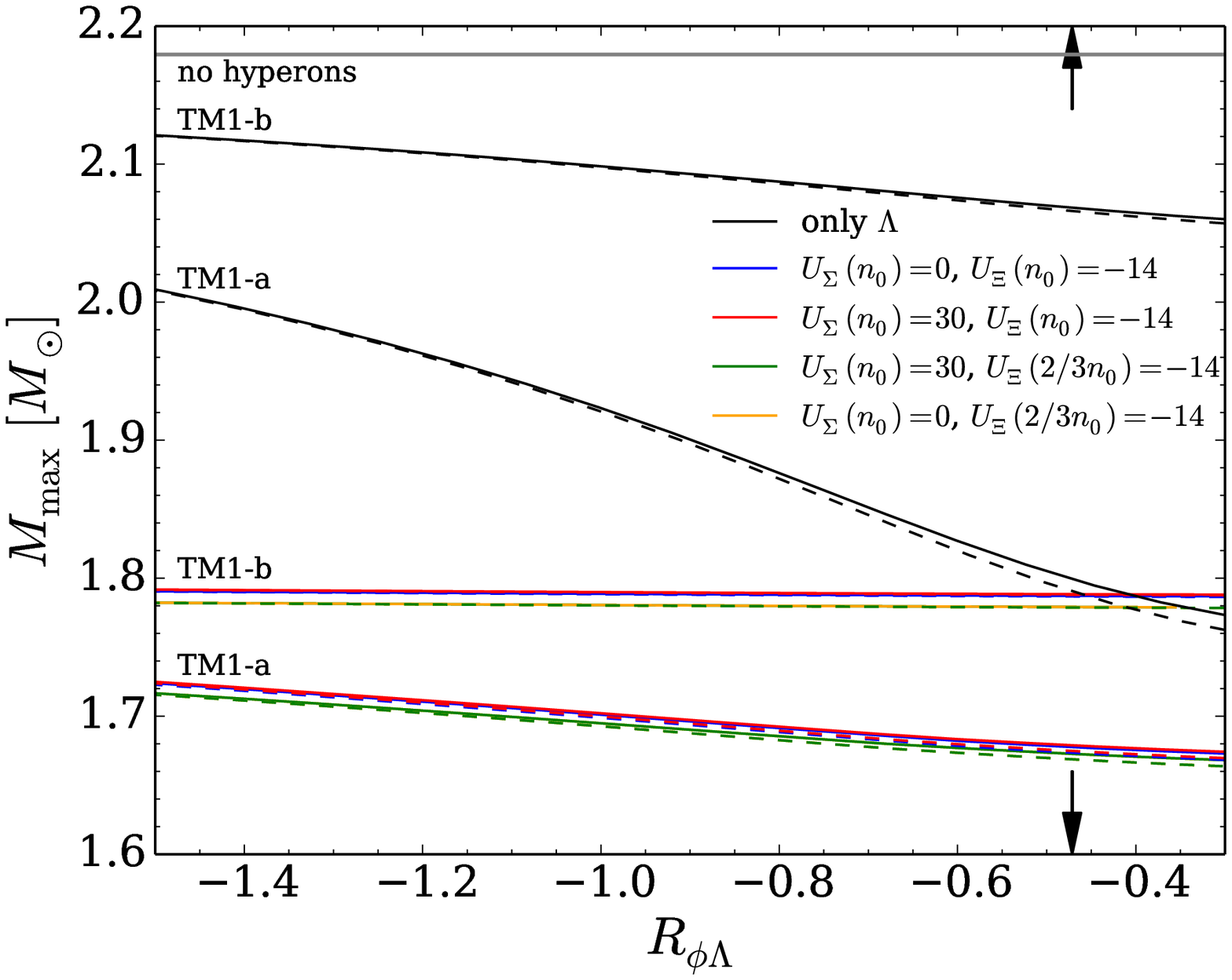}
	\includegraphics[width=\columnwidth]{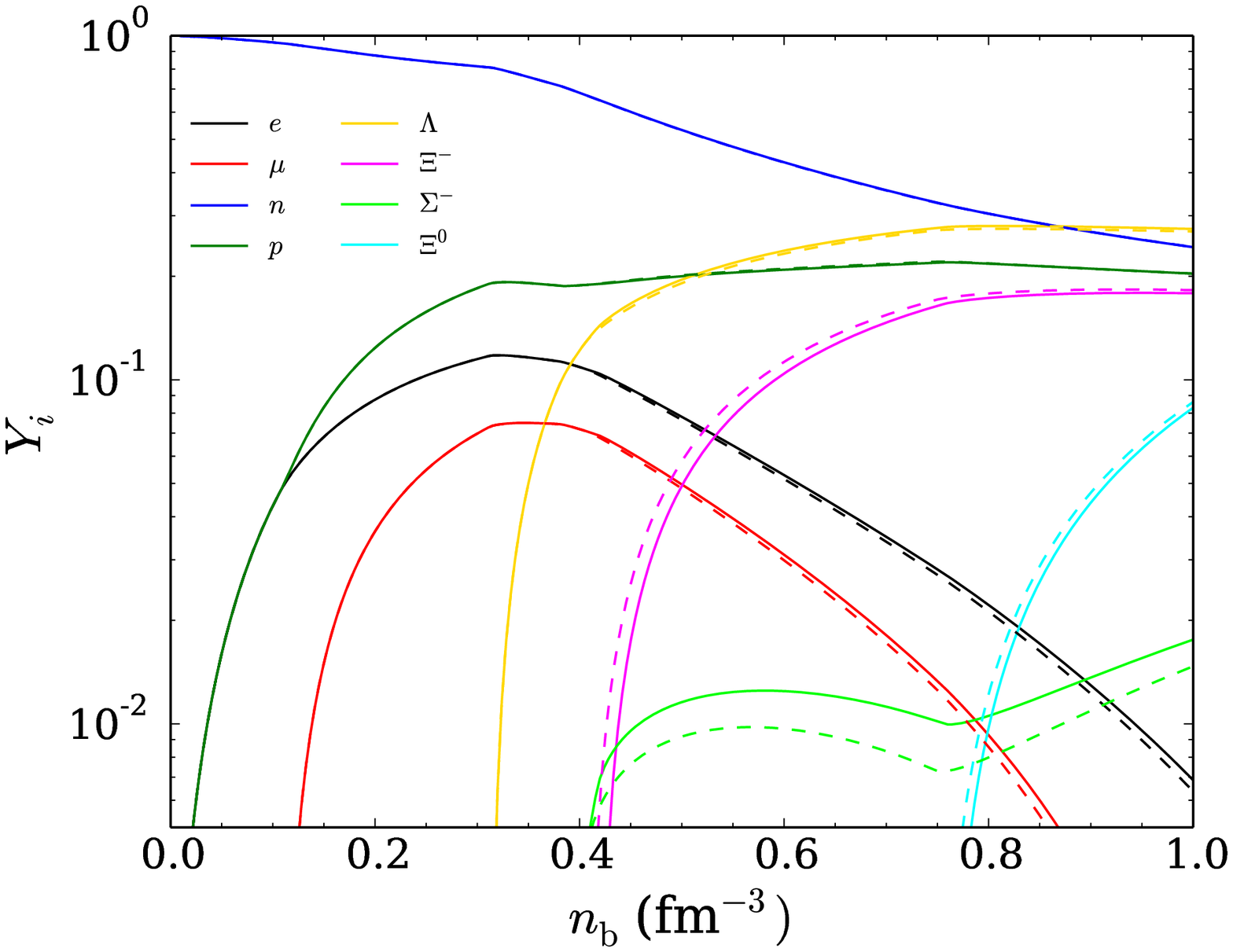}

    \caption{Left: for the TM1-a and -b models, neutron star maximum mass $M_{\rm max}$ as a function of $R_{\phi\Lambda}$ for various hyperonic models. The values $R_{\sigma\Lambda}$, $R_{\phi\Lambda}$ and $R_{\sigma^{\ast}\Lambda}$ are adjusted to reproduce the binding energies of single $\Lambda$-hypernuclei and of $_{\Lambda \Lambda}^6$He with $\Delta B_{\Lambda \Lambda}=0.50$ MeV (solid lines) and 0.84 MeV (dashed lines). The dotted line indicates the SU(6) value of $R_{\phi\Lambda}$. Right: for the TM1-a model and $R_{\phi\Lambda}$ equal to its SU(6) value and $R_{\sigma\Lambda}$, $R_{\sigma^{\ast}\Lambda}$ calibrated to hypernuclei data with $\Delta B_{\Lambda \Lambda}=0.50$ MeV, composition of the neutron star core for $U_\Sigma(n_0)=0$ MeV and  $U_\Xi(n_0)=-14$ (solid lines) or $U_\Xi(2/3n_0)=-14$ MeV (dashed lines).}
    \label{fig:TM1:2msun}
\end{figure*}

\begin{figure}
	\includegraphics[width=\columnwidth]{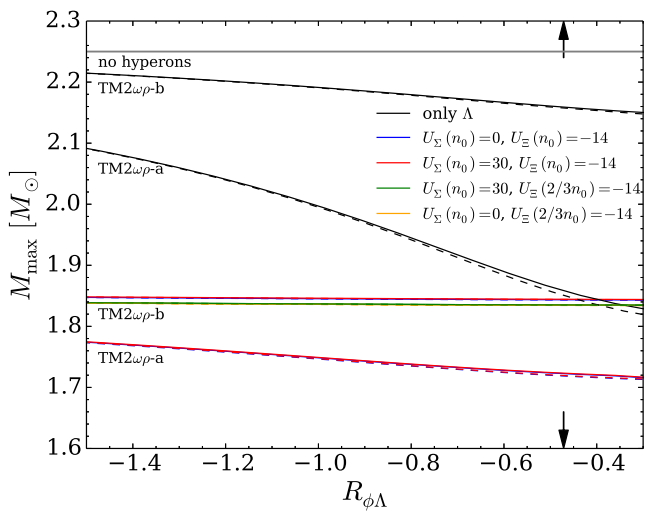}
    \caption{Analogue of Figure \ref{fig:TM1:2msun} for the TM2$\omega\rho$ parametrization.}
    \label{fig:TM2:2msun}
\end{figure}

The left panel of Figure \ref{fig:TM1:2msun} shows the maximum mass $M_{\rm max}$ obtained when solving the Tolman-Oppenheimer-Volkoff (TOV) equations \cite{tov} for an EoS based on the TM1-a parametrization as a function of $R_{\phi\Lambda}$  for the two types of hyperonic models mentioned before, both with $\Lambda$ couplings adjusted to single and double $\Lambda$-hypernuclei. In addition the horizontal grey line indicates the maximum mass obtained for a purely nucleonic core $M_{\rm max}^{\rm N}$ (see Table \ref{tab:parameters}) and the arrow shows the value of $R_{\phi\Lambda}$ corresponding to SU(6) symmetry. The right panel shows, for the TM1-a model and the two values of the $U_\Xi^N$ potential, the composition inside the core of a neutron star with a mass equal to $M_{\rm max}$, taking $R_{\phi\Lambda}$ equal to its SU(6) value and $\Lambda$ couplings adjusted to $\Delta B_{\Lambda \Lambda}=0.50$~MeV and to single hypernuclei.

\begin{figure*}
	\includegraphics[width=\columnwidth]{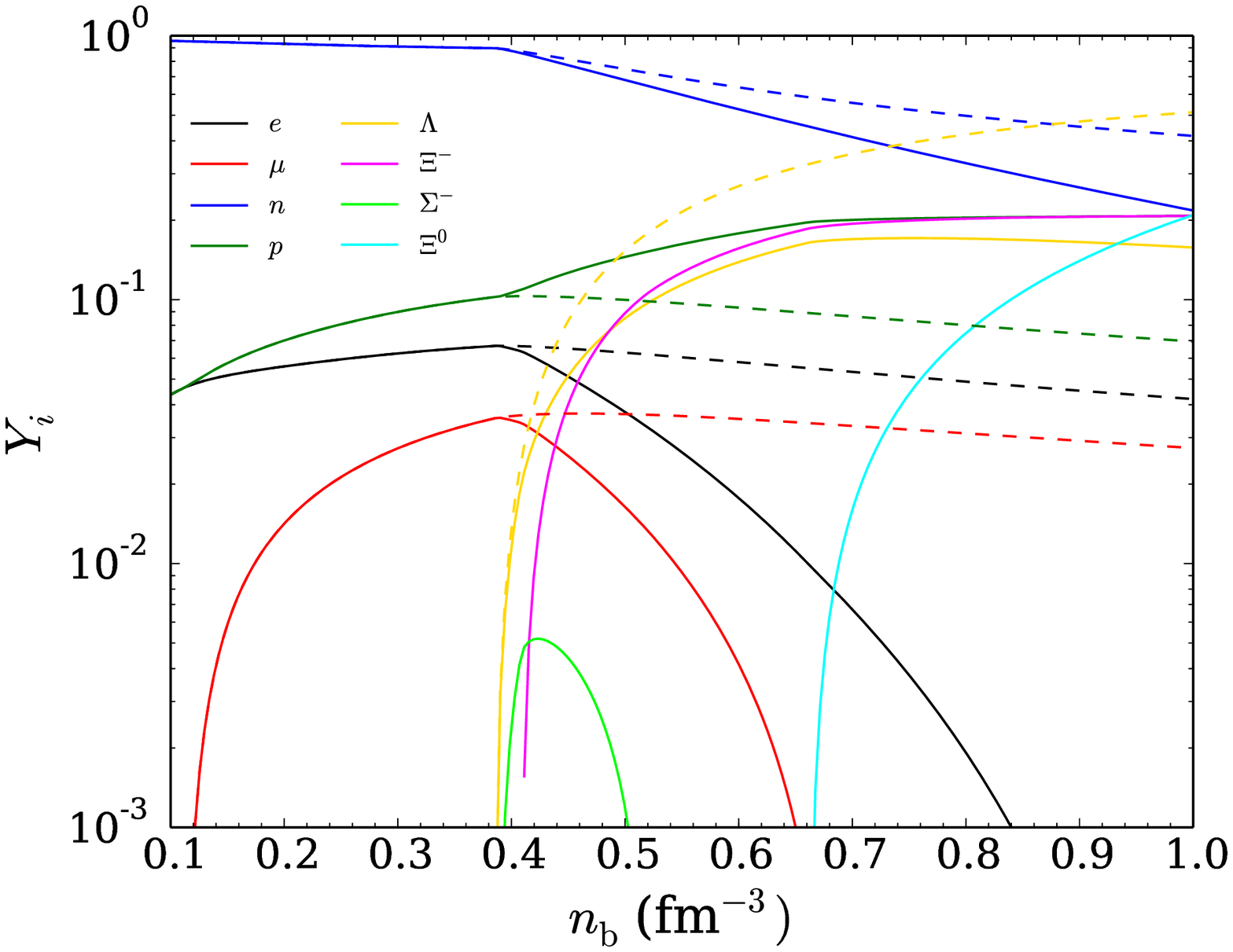}
	\includegraphics[width=\columnwidth]{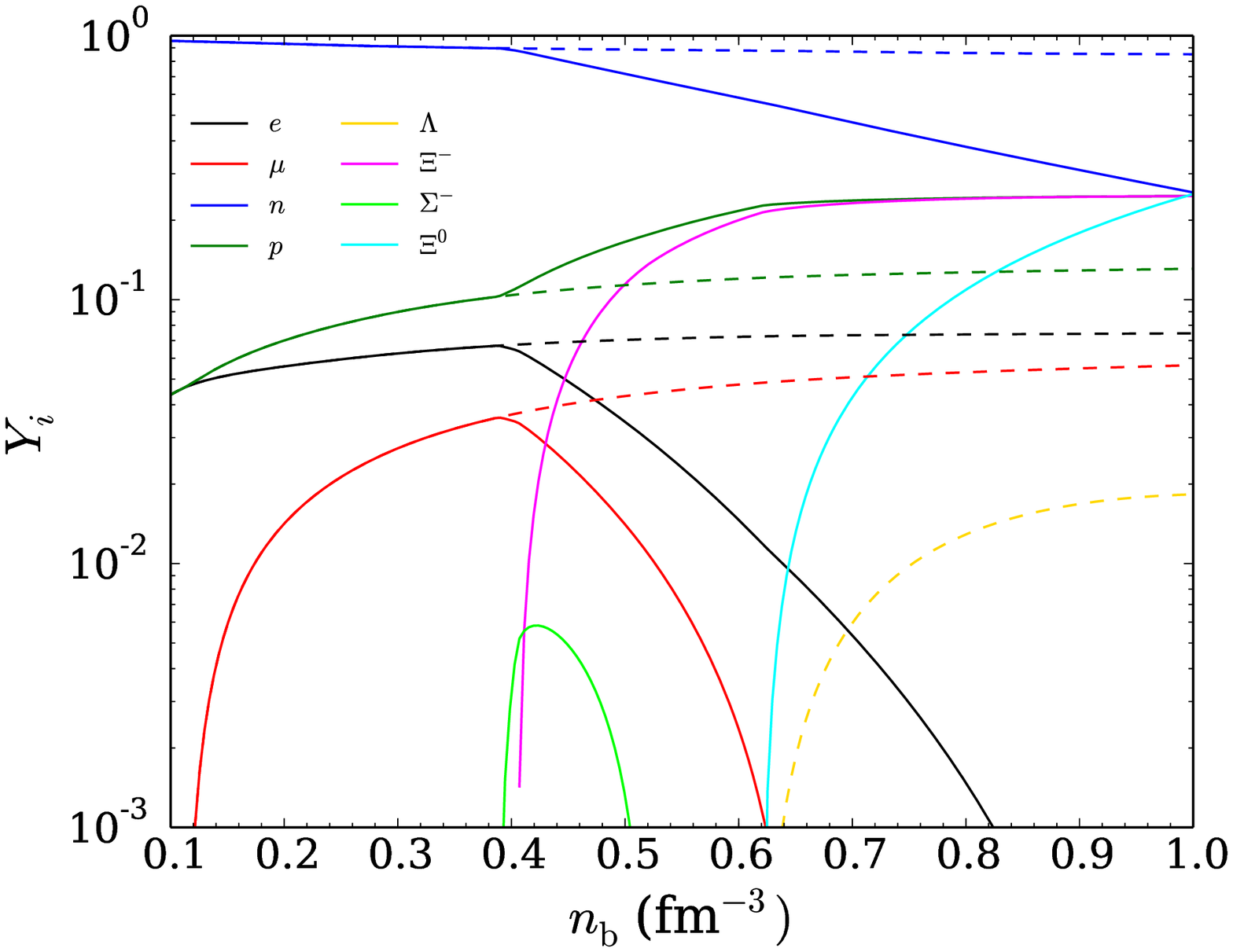}
\caption{DDME2 parametrization. Left: SU(6) i.e.. $R_{\omega\Lambda}=2/3$, right: $R_{\omega\Lambda}=1$. Obtained for $R_{\phi\Lambda}$ equal to its SU(6) value and $R_{\sigma\Lambda}$ and $R_{\sigma^*\Lambda}$ calibrated to hypernuclear data with $\Delta B_{\Lambda \Lambda}=0.5$ MeV. Dashed lines: only $\Lambda$s included, solid lines: all hyperons included with $U_\Sigma=0$ and $U_\Xi(n_0)=-14$ MeV.}
\label{fig:DDME2:compo}
\end{figure*}

The influence of the value of the potential for the $\Sigma$ hyperons on the maximum mass is found to be small. Indeed the $\Xi$ are, after the $\Lambda$, the most numerous hyperons,  owing to the fact that the $\Sigma$ potential is repulsive, and the fraction of $\Sigma$, even if they appear, is approximately one order of magnitude smaller as shown in  the  right panel of Figure \ref{fig:TM1:2msun}. Similarly, the value of the bound energy of $_{\Lambda \Lambda}^6$He hardly affects the results since very similar values of $R_{\sigma^\ast\Lambda}$ are obtained for $\Delta B_{\Lambda \Lambda}=0.50$ or 0.84 MeV as indicated in Table \ref{tab:double}. 
The lower bound for TM1-b is very flat, showing no dependence on $R_{\phi\Lambda}$ because the $\Lambda$-hyperons are suppressed and if present they only appear in residual quantities. As an example see the right panel of Fig. \ref{fig:DDME2:compo} where a similar choice of couplings for $\Lambda$ is considered. 

The second type of hyperonic models constitutes a ``maximal hyperonic model'' and sets a lower limit on the neutron star maximum mass  with an hyperonic EoS, since for the $\Sigma$ and $\Xi$ hyperons the inclusion of the vector $\phi$ meson will bring extra repulsion  even if the scalar $\sigma^*$ meson is also included, due to the vector dominance at high densities.
Consequently with the two types of hyperonic models we can calculate the range of neutron star maximum masses consistent with the available experimental data on hypernuclei and confront it to the astrophysical constraints on $M_{\rm max}$. The width of this $M_{\rm max}$ range reflects our current uncertainty or lack of information on the YN and YY interactions.  

For the TM1-a model, as shown in Figure \ref{fig:TM1:2msun}, $2\,\msun$ can only be reached when $R_{\phi\Lambda}<-1.5$, {\it i.e.,} when the SU(6) is very strongly broken, and with the condition that only $\Lambda$ hyperons are included in the model. Therefore, this model appears to be difficult to reconcile with both astrophysical and hypernuclear data. On the opposite, as far as the TM1-b model is concerned, for any value of $R_{\phi\Lambda}$, the maximal hyperonic model gives $M_{\rm max}>1.78\,\msun$ and the minimum one $M_{\rm max}<2.06-2.12\,\msun$. Thus hyperonic EoS consistent with a maximum mass of $2\,\msun$ and current hypernuclear data can be obtained.

\begin{figure*}
	\includegraphics[width=\columnwidth]{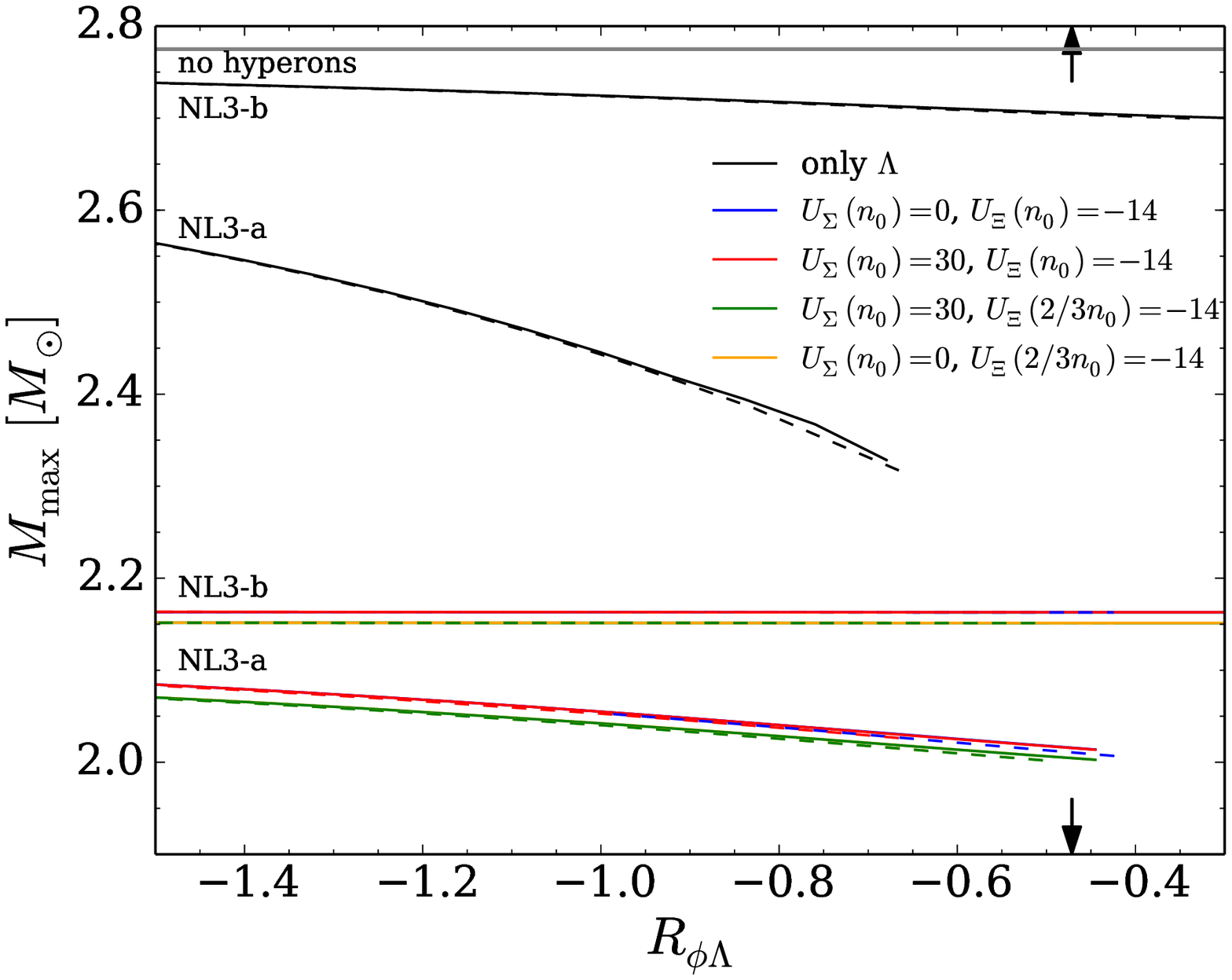}
	\includegraphics[width=\columnwidth]{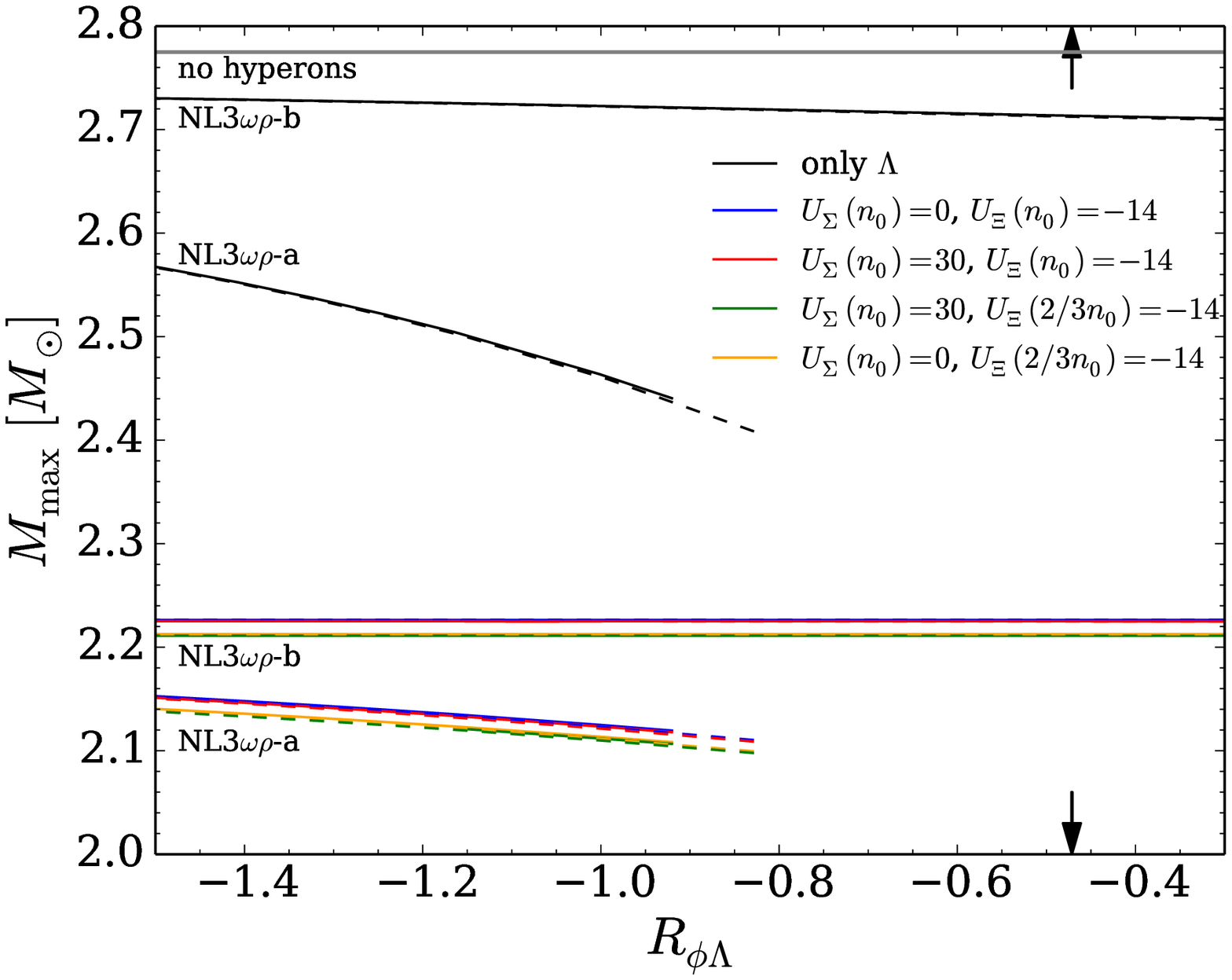}
    \caption{Analogue of Figure \ref{fig:TM1:2msun} for the NL3 (left) and NL3$\omega\rho$ (right) parametrizations.}
    \label{fig:NL3:2msun}
\end{figure*}
 A similar approach is used for the four additional parametrizations. As shown in Figure \ref{fig:TM2:2msun} for TM2$\omega\rho$-a model a maximum mass of $2\,\msun$ is reached if $R_{\sigma^{\ast}\Lambda}\leq -1$ for the minimal hyperonic model. Again, as for TM1, the breaking SU(6) symmetry is required, but to a lesser extent since maximum masses stars are larger for the TM2$\omega\rho$ parametrization than for TM1 one. For the TM2$\omega\rho$-b model, the maximum mass reachable for the minimal hyperonic model is always larger than $2\,\msun$: $M_{\rm max}>2.15-2.22\,\msun$. This model is thus compatible with both hypernuclear and astrophysical data.

Very similar results are obtained when comparing the NL3 and NL3$\omega\rho$ parametrizations, shown in Figure \ref{fig:NL3:2msun}. It has to be mentioned that for models -a and for $R_{\phi\Lambda}$ small in absolute value, 
the maximum value of the mass can not be obtained when the function of the mass in terms of the density starts decreasing, which is the very definition of the maximum mass, but when the baryon effective mass becomes equal to 0. Such cases are not plotted in Figure\,\ref{fig:NL3:2msun}. Any model for the NL3 and NL3$\omega\rho$ parametrizations is consistent with the existence of a $2\,\msun$ neutron star and they even predict the possibility of having hyperonic neutron stars with masses at least larger than $2.1\,\msun$.

\begin{figure*}
	\includegraphics[width=\columnwidth]{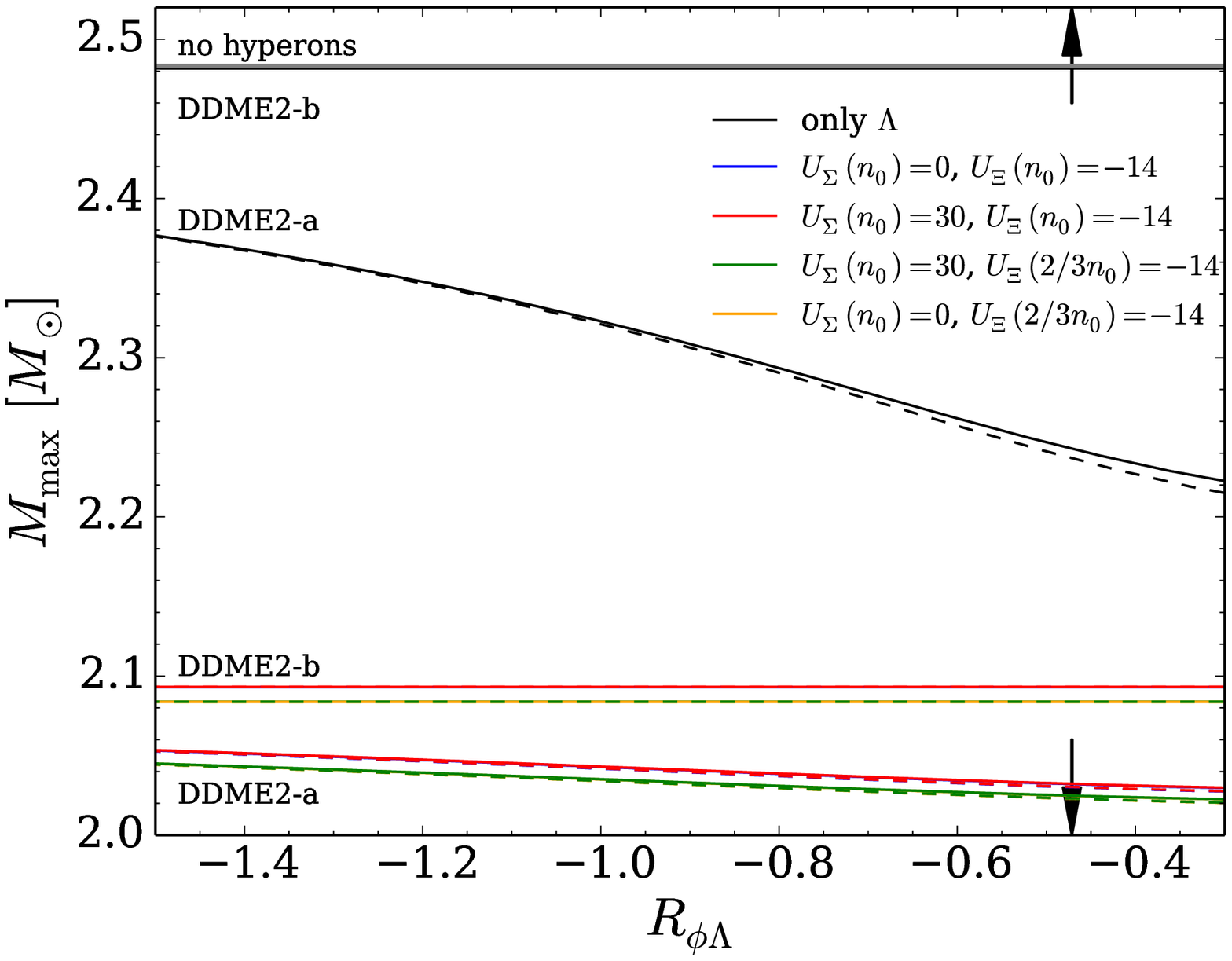}
	\includegraphics[width=\columnwidth]{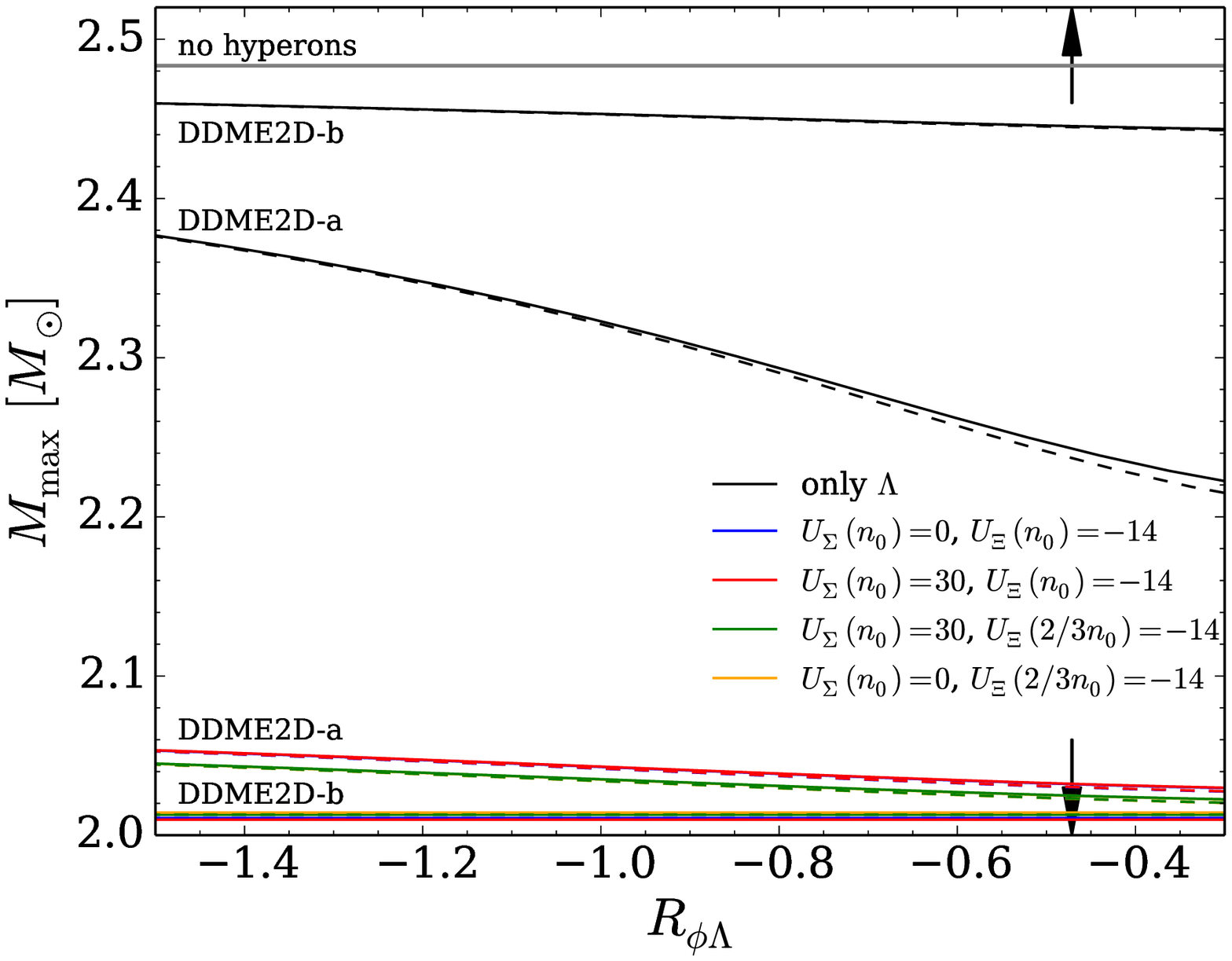}
    \caption{Analogue of Figure \ref{fig:TM1:2msun} for the DDME2 (left) and DDME2D (right) parametrizations.}
   \label{fig:DDME2:2msun}
\end{figure*}

 Results for the DDME2 and DDME2D parametrizations are plotted in Figure \ref{fig:DDME2:2msun}. Whether hyperon couplings are density-dependent or not appears not to affect the maximum mass for the models a, while it does for the models b.

 For the DDME2 parametrization, Figure \ref{fig:DDME2:compo} shows that for the model a the
 $\Lambda$ is the first hyperon to set in. For a density slightly
larger the $\Sigma^-$ also appears and the fraction of electrons and
$\mu$ decreases. However, as soon the $\Xi^-$ sets in it is favoured
because the repulsive $g_{\omega\Xi}$ coupling is one half of the coupling
of the $\omega$ to the $\Sigma^-$. Taking $g_{\omega\Lambda}=1$, the
$\Lambda$ hyperon  becomes disfavoured and only a small fraction at quite
high densities appears in the case with only nucleons and
$\Lambda$'s. If the other hyperons are also taken into account there
is no $\Lambda$'s below $n_{\rm b}=1$ fm$^{-3}$.

\begin{figure*}
	\includegraphics[width=\columnwidth]{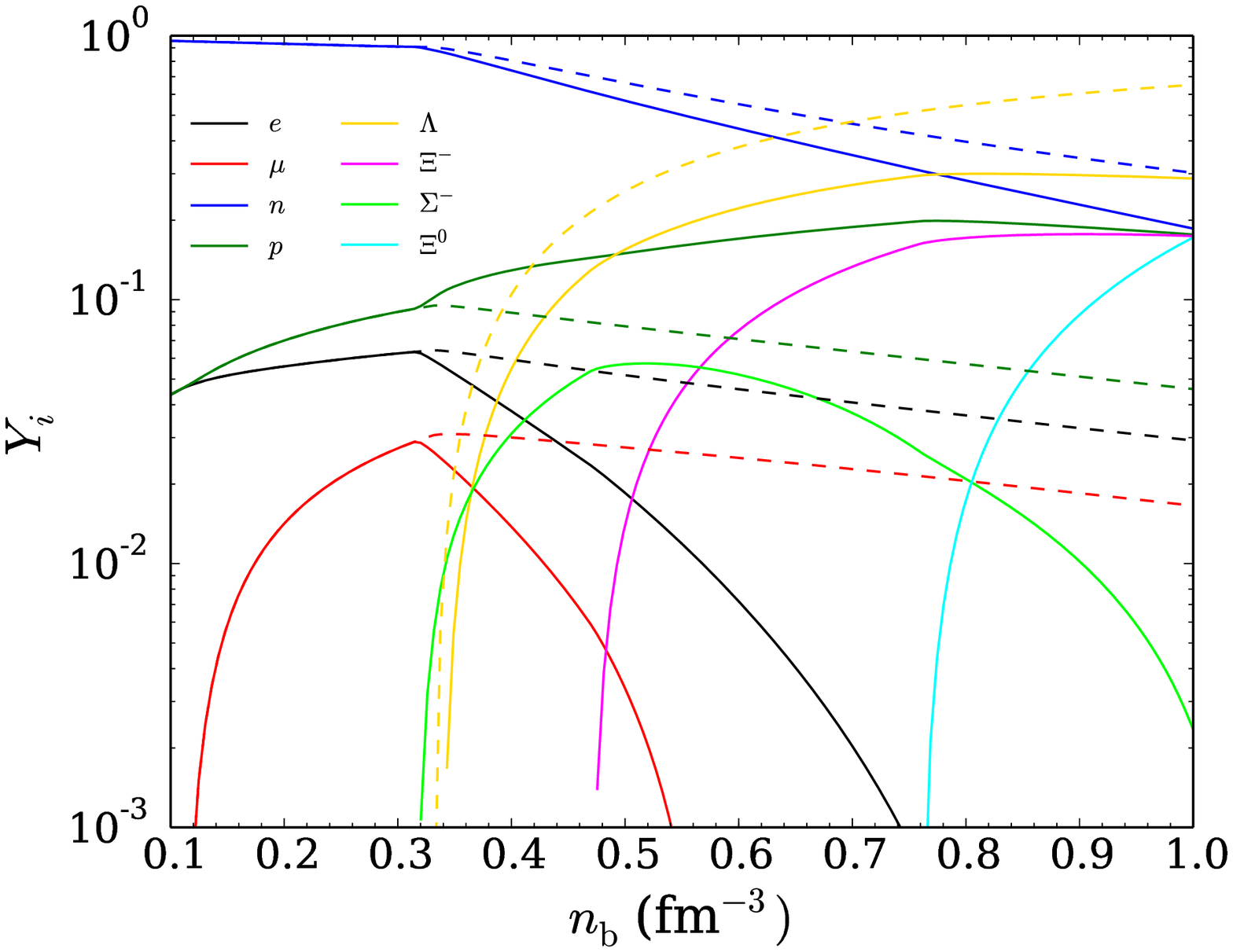}
	\includegraphics[width=\columnwidth]{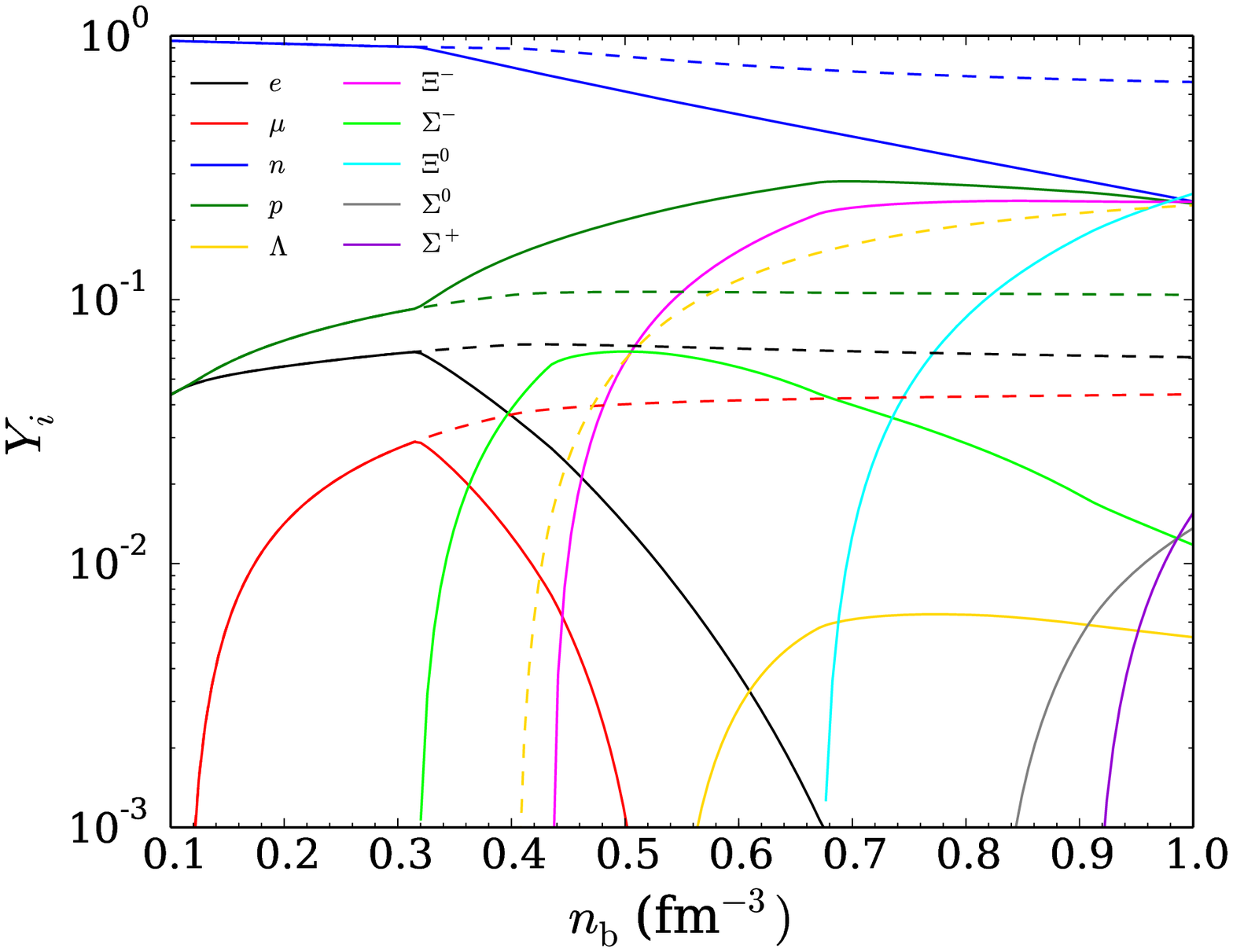}
\caption{Same as Figure \ref{fig:DDME2:compo} for DDME2D parametrization.}
\label{fig:DDME2D:compo}
\end{figure*}

 For the DDME2D parametrization the  hyperon-meson couplings are weaker than in the
previous scenario, and they decrease with the density as the
nucleon-meson couplings. A weaker $g_{\omega\Sigma}$ allows a lower onset
density, and since in this model the $\rho$-meson coupling is quite
strong the $\Sigma^{-}$ sets in first, as shown in Figure \ref{fig:DDME2D:compo}.
 The $\Lambda$ hyperon sets in at a density
very close to $\Sigma^{-}$ if $R_{\omega\Lambda}=2/3$, otherwise if
$R_{\omega\Lambda}=1$ its onset is shifted to quite high densities and
its fraction is always below 1\%. As in the case with constant
couplings, as soon as the $\Xi^-$ sets in the amount of the $\Sigma^{-}$
decreases steadily, since $R_{\omega\Xi}=1/3$ is half the
corresponding coupling for the $\Sigma^{-}$. Having more strict
constraints to fix the different hyperon-meson couplings and
including the strangeness hidden mesons, $\sigma^*$ and $\phi$, the
relative abundances will certainly change, but the total amount of
strangeness is less sensitive to the relative magnitude of the
couplings.  For instance, making the $\Sigma$ potential in nuclear
matter repulsive will certainly reduce the amount of $\Sigma^{-}$
present in matter increasing the amount of the other hyperons.

\section{Summary and Conclusions}
\label{sect:conc}
 Modelling single and double $\Lambda$-hypernuclei  we first calibrate the $\Lambda$ couplings for six different RMF parametrizations. The usual way of calibrating the $R_{\sigma\Lambda}$ coupling by imposing the value of the $\Lambda$ potential in symmetric baryonic matter, {\it i.e.,} using  $U_\Lambda^N(n_0)\simeq-30$ MeV, appears in agreement with the binding energy of single $\Lambda$-hypernuclei in the s- and p-shells. Moreover, the value of $R_{\sigma\Lambda}$ that comes out of the order of $\sim 0.62$ when the SU(6) value for $g_{\omega\Lambda}$ is taken is quite independent of the model considered.
This is not at all the case for the calibration of the $R_{\sigma^\ast\Lambda}$ and $R_{\phi\Lambda}$ couplings. Calibrating to the bound energy $_{\Lambda \Lambda}^6$He shows that for the models a and b the  $\Lambda$ potential in pure $\Lambda$ matter $U_\Lambda^\Lambda(n_0)$ varies between $\sim -17$ and $\sim +9$MeV. This is at variance with the usual values of $-1$ or $-5$ MeV employed in the literature, showing that these values are inconsistent with the hypernuclear data.  Tables \ref{tab:single} and \ref{tab:single} provide the values of the various couplings to the $\Lambda$ calibrated to hypernuclear data.

We then proceed by constructing unified hyperonic EoS for neutron star matter. While an approach similar to the one presented for the $\Lambda$ hyperon should in principle be used for the $\Sigma$ and $\Xi$ ones,  the lack of hypernuclear experimental data does not allow us to calibrate their couplings to hypernuclei properties. Consequently, we proceed by devising two limiting hyperonic models. In the minimal one only the $\Lambda$ hyperon is included in addition to the nucleons and its couplings calibrated to hypernuclear data. The $\Sigma$  and $\Xi$ hyperons are included in the maximal hyperonic model using a repulsive potential in symmetric baryonic matter for the $\Sigma$ and a value for $\Xi$ of $-14$~MeV consistent with the scarce experimental constraint for this hyperon. For these two hyperons no coupling to the hidden mesons $\sigma^*$ and $\phi$ is included because of the non-existing experimental data that would allow to constrain the coupling parameters.

Finally, we confront the EoS calibrated to hypernuclear data to the astrophysical constraint that neutron stars with $2\,\msun$ exist. For the TM1 and TM2$\omega\rho$ parametrizations the breaking of the SU(6) symmetry appears required to be consistent with this constraint, and it is still not clear if even breaking the SU(6) the 2 $M_\odot$ limit is satisfied when all the hyperons of the baryonic octet also interact with the $\sigma^*$ and  $\phi$ mesons. The NL3, NL3$\omega\rho$ models predict the existence of hyperonic stars with masses larger than at least $2\,\msun$.  However, these two models pose some problems due to the fact that  the effective mass becomes negative at densities lower than that of the maximum mass for values of the $R_{\phi\Lambda}$ coupling close to the SU(6) value, meaning that the models are unacceptable if future constraints indicate that the appropriate couplings lie in the range that results in a negative effective mass. The DDME2 and DDME2D models are both consistent with the $2\,\msun$ constraint.

In conclusion it still appears difficult to exclude any of the parametrizations used in this work on the ground that hyperonic stars are not consistent with the existence of $2\,\msun$ although our models are calibrated to up-to-date hypernuclei data. This reflects the fact that the properties of the nucleonic sector are themselves hardly constrained  at high density. Future measurements of neutron star properties (mass and radius, surface gravitational redshift, moment of inertia, \ldots)  and of high density properties of asymmetric nuclear matter in the laboratory  appear necessary to constrain further the nucleonic EoS. If in addition properties of $\Xi$ and $\Sigma$ hyperons are better constrained   one could reduce the range of possible maximum masses given by the minimal and maximal hyperonic models, and potentially solve the hyperon puzzle.

\begin{acknowledgments}
Partial support comes from ``NewCompStar'', COST Action MP1304. The work of M.F. has been partially supported by the NCN (Poland) Grant No. 2014/13/B/ST9/02621 and by the  STSM grant from the  COST Action MP1304, and by  Funda\c c\~ao para a Ci\^encia e
Tecnologia (FCT), Portugal, under the project No. UID/FIS/04564/2016.
\end{acknowledgments}

\newpage{\pagestyle{empty}\cleardoublepage}


\begin{thebibliography}{100}

\bibitem{ambart} V. A. Ambartsumyan and G. S. Saakyan, Sov. Astron. {\bf 4}, 187 (1960).

\bibitem{micro} H.-J. Schulze, M. Baldo, U. Lombardo, J. Cugnon and A. Lejeune, Phys. Lett. B {\bf 355} 21 (1995);
                         H.-J. Schulze, M. Baldo, U. Lombardo, J. Cugnon and A. Lejeune, Phys. Rev. C {\bf 57}, 704 (1998); 
                         M. Baldo, G. F. Burgio and H.-J. Schulze, Phys. Rev. C {\bf 58}, 3688 (1998).
                         M. Baldo, G. F. Burgio and H.-J. Schulze, Phys. Rev. C {\bf 61}, 055801 (2000);
                         I. Vida\~na, A. Polls, A. Ramos, M. Hjorth-Jensen and V. G. J. Stoks, Phys. Rev. C {\bf 61}, 025802 (2000);
                         I. Vida\~na, A. Polls, A. Ramos, L. Engvik and M. Hjorth-Jensen, Phys. Rev. C {\bf 62}, 035801 (2000);
                         H-J. Schulze, A. Polls, A. Ramos and I. Vida\~na, Phys. Rev. C {\bf  73}, 058801 (2006);
                         H.-J. Shulze and T. Rijken, Phys. Rev. C {\bf 84}, 035801 (2011).

\bibitem{vlowk} H. Dapo, B.-J. Schaefer and J. Wambach, Phys. Rev. C {\bf 81}, 035803 (2010).

\bibitem{dbhf1} F. Sammarruca, Phys. Rev. C {\bf 79} 034301 (2009).

\bibitem{dbhf2} T. Katayama and K. Saito, arXiv:1410.7166 (2014);
                          arXiv:1501.05419 (2015).

\bibitem{qmc} D. Lonardoni, F. Pederiva and S. Gandolfi, Phys. Rev. C {\bf 89} 014314 (2014).

\bibitem{rmf} N. K. Glendenning, Phys. Lett. B {\bf 114}, 392 (1982); 
                                   N. K. Glendenning, Astrophys. J. {\bf 293}, 470 (1985);
                                   N. K. Glendenning, Z. Phys. A {\bf 326}, 57 (1987); 
                                   N. K. Glendening and S. A. Moszkowski, Phys. Rev. Lett. {\bf 67}, 2414 (1991);
                                  F. Weber and M. K. Weigel, Nucl. Phys. A {\bf 505}, 779 (1989). 
                                  R. Knorren, M. Prakash and P. J. Ellis, Phys. Rev. C {\bf 52}, 3470 (1995);
                                  J. Schaffner and I. Mishustin, Phys. Rev. C {\bf 53}, 1416 (1996);
                                 H. Huber, F. Weber, M. K. Weigel and Ch. Schaab, Int. J. Mod. Phys. E {\bf 7}, 310 (1998).

\bibitem{shf} S. Balberg and A. Gal, Nucl. Phys. A {\bf 625}, 435 (1997);
                     S. Balberg, I. Lichtenstadt and G. B. Cook, Astrophys. J. Suppl. Ser. {\bf 121}, 515 (1999).
                     D. E. Lanskoy and Y. Yamamoto, Phys. Rev. C {\bf 55}, 2330 (1997);
                     T. Y. Tretyakova and D. E. Lanskoy, Eur. Phys. J. A {\bf 5}, 391 (1999);
                     J. Cugnon, A. Lejeune and H.-J. Schulze, Phys. Rev. C {\bf 62}, 064308 (2000);
                     I. Vida\~na, A. Polls, A. Ramos and H.-J. Schulze, Phys. Rev. C {\bf 64}, 044301 (2001);
                     X.-R. Zhou , H.-J. Schulze, H. Sagawa, C.-X. Wu and E.-G. Zhao, Phys. Rev. C {\bf 76}, 034312 (2007);
                     X.-R. Zhou, A. Polls, H.-J. Schulze and I. Vida\~na, Phys. Rev. C {\bf 78}, 054306 (2008).

\bibitem{hulsetaylor} R. A. Hulse and J. H. Taylor, Astrophys. J. Lett. {\bf 195} L51 (1975).

%

\bibitem{bombaci08} I. Bombaci, P. K. Panda, C. Provid\^encia, and Isaac Vida\~na, Phys. Rev. D 77, 083002  (2008)

\bibitem{cavagnoli11} R. Cavagnoli, D. P.Menezes and C. Provid\^encia,  Phys. Rev. C 84,  065810 (2011)

\bibitem{demorest} P. Demorest {\it et al.,} Nature {\bf 467}, 1081 (2010).

\bibitem{fonseca} Fonseca, E., Pennucci, T.~T., Ellis, J.~A., et al.\ 2016, arXiv:1603.00545 

%

\bibitem{antoniadis} J. Antoniadis {\it et al.,} Science {\bf 340} 6131 (2013).

\bibitem{fortin2015} M. Fortin, J.~L. Zdunik, P. Haensel and M. Bejger, Astron. and Astrophys. {\bf 576}, A68 (2015).


\bibitem{Bednarek11} I. Bednarek, P. Haensel, J. L. Zdunik, M. Bejger and R. Ma\'nka, Astron. and Astrophys. {\bf 543}, A157 (2012).

\bibitem{Weissenborn} S. Weissenborn, D. Chatterjee, and J. Schaffner--Bielich, Phys. Rev. C {\bf 85}, 065802 (2012).

\bibitem{Oertel14} M. Oertel, C. Provid\^encia, F. Gulminelli and Ad. R. Raduta,  J. Phys. G. {\bf 42} 075202 (2015).

\bibitem{Maslov} K. A. Maslov, E. E. Kolomeitsev and D. N. Voskresensky, Phys. Lett. B {\bf 748} 369 (2015).

\bibitem{Dalen} E. N. E. van Dalen, G. Colucci and A. Sedrakian, Phys. Lett. B {\bf 734}, 383 (2014).

\bibitem{Gandolfi12} S. Gandolfi, J. Carlson and S. Reddy, Phys. Rev. C {\bf 85}, 032801 (2012).

\bibitem{Hebeler13} K. Hebeler, J.~M. Lattimer, C.~J. Pethick, and A.  Schwenk, ApJ {\bf 773}, 11 (2013).

\bibitem{taka} T. Takatsuka {\it et al.,} Eur. Phys. J. A {\bf 13}, 213 (2002); Prog. Theor. Phys. Suppl. {\bf 174}, 80 (2008).

\bibitem{vidanatbf} I. Vida\~na, D. Logoteta, C. Provid\^encia, A. Polls and I. Bombaci, Eur. Phys. Lett. {\bf 94}, 11002 (2011).

\bibitem{yamamoto} Y. Yamamoto, T. Furumoto, B. Yasutake and Th. A. Rijken, Phys. Rev. C {\bf 88} 022801 (2013); Phys. Rev. C {\bf 90} 045805 (2014).

\bibitem{lonardoniprl} D. Lonardoni, A. Lovato, S. Gandolfi and F. Pederiva Phys. Rev. Lett. {\bf 114} 092301 (2015).

\bibitem{Ozel} F. \"Ozel, D. Psaltis, S. Ransom, P. Demorest and M. Alford, Astrophys. J. Lett. {\bf 724}  L199 (2010).

\bibitem{WeissenbornSagert} S. Weissenborn, I. Sagert, G. Pagliara, M. Hempel and J. Schaeffner-Bielich, Astophys. J. Lett. {\bf 740} L14 (2011).

\bibitem{Klahn2013} T. Kl\"ahn, D. Blaschke and  D. Lastowiecki, Phys. Rev. D {\bf 88} 085001 (2013).

\bibitem{Bonanno} L. Bonanno and A. Sedrakian, Astron. and Astrophys. {\bf 539} 416 (2012).

\bibitem{Lastowiecki2012} R. Lastowiecki, D. Blaschke, H. Grigorian and S. Typel Acta Phys. Polon. Suppl., {\bf 5} 535 (2012).

\bibitem{Drago} A. Drago, A. Lavagno, G. Pagliara and D. Pigato, Phys. Rev. C {\bf 90}, 065809 (2014); EPJ Web Conf. {\bf 95}, 01011 (2015).


\bibitem{haidenbauer16} J. Haidenbauer, U.-G. Meissner, N. Kaiser and W. Weise, arXiv:1621.03758v1 (2016).

\bibitem{Miller2016} M.~C.~Miller and F.~K.~Lamb, Eur.\ Phys.\ J.\ A 52, no. 3, 63 (2016)

\bibitem{Haensel2016} Haensel, P., Bejger, M., Fortin, M., Zdunik, J.L. \ 2016, Eur. Phys. J. A 52, 59

\bibitem{nicer} K.~C. Gendreau,  Z. Arzoumanian and T. Okajima, Proc. SPIE, 8443 (2012).

\bibitem{athena} C. Motch, J. Wilms, D. Barret {\em et al.} , arXiv:1306.2334 (2013).

\bibitem{loft} M. Feroci, J.~W. den Herder, E. Bozzo {\em et al.} Proc. SPIE 8443 (2012).

\bibitem{Shen06} H. Shen, F. Yang, H. Toki, Prog. Theor. Phys. 115, 325 (2006).

\bibitem{Sedrakian14} van Dalen, E.~N.~E., Colucci, G., \& Sedrakian, A.\ 2014, Physics Letters B, 734, 383

\bibitem{TM1} Y. Sugahara, and H. Toki, Nucl. Phys. {\bf A, 579}, 557 (1994).

\bibitem{providencia13} C. Provid\^encia and Aziz Rabhi, Phys. Rev. C 87, 055801 (2013).

\bibitem{NL3} G.~A. Lalazissis, J. K{\"o}nig, and P. Ring, Phys. Rev. C {\bf 55}, 540 (1997).

\bibitem{NL3wra} C.~J. Horowitz, and J. Piekarewicz,  Phys. Rev. Lett. {\bf 86}, 5647 (2001).

\bibitem{DDME2} G.~A. Lalazissis, T. Nik{\v s}i{\'c}, D. Vretenar, and P. Ring, Phys. Rev. C {\bf 71}, 024312 (2005).

\bibitem{stosa} H. Shen, H. Toki, K. Oyamatsu, and K. Sumiyoshi, Nucl. Phys. A 637, 435 (1998).

\bibitem{stosb} H. Shen, H. Toki, K. Oyamatsu, and K. Sumiyoshi, Astrophys. J. Suppl. 197, 20 (2011).

\bibitem{Fortin16} Fortin, M., Provid{\^e}ncia, C., Raduta, A.~R., et al.\ 2016, \prc, 94, 035804

\bibitem{danielewicz02} P. Danielewicz, R. Lacey, W.G. Lynch, Science 298, 1592 (2002).

\bibitem{Dutra14} M. Dutra, O. Louren{\c c}o, S.~S. Avancini, {\em et al.}, Phys. Rev. C {\bf 90}, 055203 (2014).

\bibitem{gm91} N. K. Glendenning and S. A. Moszkowski,  { Phys. Rev. Lett.} {\bf 67},  2414 (1991).

\bibitem{Schaffner96} J.  Schaffner  and I. N.  Mishustin,  {Phys. Rev.} C {\bf 53},  1416 (1996)

\bibitem{engelmann66} R. Engelmann {\it et al.,} Phys. Lett. {\bf 21}, 587 (1966).

\bibitem{alexander68} G. Alexander {\it et al.,} Phys. Rev. {\bf 173}, 1452 (1968)

\bibitem{sechi68} B. Sechi--Zor {\it et al.,} Phys. Rev. {\bf 175}, 1735 (1968).

\bibitem{kadyk71} J. A. Kadyk {\it et al.,} Nucl. Phys. B {\bf 27}, 13 (1971).

\bibitem{eisele71} J. Eisele {\it et al.,} Phys. Lett. B {\bf 37}, 204 (1971).

\bibitem{dapnie52} M. Danysz and J. Pniewski, Phil. Mag. {\bf 44}, 348 (1953).

\bibitem{dl0} M. Danysz {\it et al.,} Phys. Rev. Lett. {\bf 11}, 29 (1963).

\bibitem{dl1} M. Danysz {\it et al.,} Nucl. Phys. A {\bf 49}, 121 (1963).

\bibitem{dl1b} R. H. Dalitz, D. H. Davis, P. H. Fowler, A. Montwill, J. Pniewski and J. A. Zakrewski, Proc. Roy. Soc. London, Ser. A {\bf 426}, 1 (1989).

\bibitem{dl2} D. J. Prowse, Phys. Rev. Lett. {\bf 17}, 782 (1966).

\bibitem{dl3} S. Aoki {\it et al.,} Prog. Theor. Phys. {\bf 85}, 1287 (1991).

\bibitem{dl4} C. B. Dover, D. J. Millener, A. Gal and D. H. Davis, Phys. Rev. C {\bf 44}, 1905 (1991).

\bibitem{dl5} G. B. Franklin, Nucl. Phys. A {\bf 585}, 83c (1995).

%

\bibitem{nagara} H. Takahashi {\it et al.,} Phys. Rev. Lett. {\bf 87}, 212502 (2001).

\bibitem{khaustov00} P. Khaustov {\it et al.,} Phys. Rev. C {\bf 61}, 054603 (2000).

\bibitem{nakazawa15} K. Nakazawa {\it et al.,} Prog. Theor. Exp. Phys., 0033D02 (2015).


\bibitem{bertini80} R. Bertini {\it et al.,} Phys. Lett. B {\bf 90}, 375 (1980).

\bibitem{bertini84} R. Bertini {\it et al.,} Phys. Lett. B {\bf 136}, 29 (1984).

\bibitem{bertini85} R. Bertini {\it et al.,} Phys. Lett. B {\bf 158}, 19 (1985).

\bibitem{piekarz82} H. Piekarz {\it et al.,} Phys. Lett. B {\bf 110}, 428 (1982).

\bibitem{yamazaki85} T. Yamazaki {\it et al.,} Phys. Rev. Lett. {\bf 54}, 102 (1985).

\bibitem{tang88} L. Tang {\it et al.,} Phys. Rev. C {\bf 38}, 846 (1988).

\bibitem{bart99} S. Bart {\it et al.,} Phys. Rev. Lett. {\bf 83}, 5238 (1999).

\bibitem{hayano89} R. Hayano {\it et al.,} Phys. Lett. B {\bf 231}, 355 (1989).

\bibitem{nagae98} T. Nagae {\it et al.,} Phys. Rev. Lett. {\bf 80}, 1605 (1998).

\bibitem{dgm89} C. B. Dover, D. J. Millener and A. Gal, Phys. Rep. {\bf 184}, 1 (1989).

\bibitem{batty1} C. J. Batty, E. Friedman and A. Gal, Phys. Lett. B {\bf 335}, 273 (1994).

\bibitem{batty2} C. J. Batty, E. Friedman and A. Gal, Prog. Theor. Phys. Suppl. {\bf 117}, 227 (1994).

\bibitem{batty3} C. J. Batty, E. Friedman and A. Gal, Phys. Rep. {\bf 287}, 385 (1997).

\bibitem{mares95} J. Mare\v{s}, E. Friedman, A. Gal and B. K. Jennings, Nucl. Phys. A {\bf 594}, 311 (1995).

\bibitem{dabrowski99} J. D\c{a}browski, Phys. Rev. C {\bf 60}, 025205 (1999).

\bibitem{noumi02} H. Noumi {\it et al.,} Phys. Rev. Lett. {\bf 89}, 072301 (2002); {\it Erratum}: Phys. Rev. Lett. {\bf 90}, 049903(E) (2003).

\bibitem{saha04} P. K. Saha {\it et al.,} Phys. Rev. C {\bf 70}, 044613 (2004).

\bibitem{harada05} T. Harada and Y. Hirabayashi, Nucl. Phys. A {\bf 759}, 143 (2005).

\bibitem{harada06} T. Harada and Y. Hirabayashi, Nucl. Phys. A {\bf 767}, 206 (2006).

\bibitem{hugenford94} E. V. Hungerford, Prog. Theor. Phys. Suppl. {\bf 117}, 135 (1994).

\bibitem{hypHI} S. Bianchin {\it et al.,} Int. J. Mod. Phys. E {\bf 18}, 2187 (2009).

\bibitem{rappold13} C. Rappold {\it et al.,} Nucl. Phys. A {\bf 913}, 170 (2013).


\bibitem{hashimoto06} O. Hashimoto and H. Tamura, Prog. Part. Nucl. Phys. {\bf 57}, (2006) 564.

\bibitem{kau} P. Khaustov {\it et al.,} Phys. Rev. C {\bf 61}, (2000) 054603.

\bibitem{nijmegen} P. M. M. Maesen, T. A. Rijken, and J. J.de  Swart, Phys. Rev. C {\bf 40}, (1989) 2226;
                                T. A. Rijken, V. G. J. Stoks, and Y. Yamamoto, Phys. Rev. C {\bf 59}, (1999) 21; 
                                V. G. J. Stoks and T. A. Rijken, Phys. Rev. C {\bf 59}, (1999) 3009;
                                T. A. Rijken, Phys. Rev. C {\bf 73}, (2006) 044007;
                                T. A. Rijken and Y. Yamamoto, Phys. Rev. C {\bf 73}, (2006) 044008.

\bibitem{Ahn13} Ahn, J.~K., Akikawa, H., Aoki, S., et al.\ 2013, \prc, 88, 014003

\bibitem{Sugahara94} Sugahara, Y., \& Toki, H.\ 1994, Progress of Theoretical Physics, 92, 803

\bibitem{Avancini07} S.S. Avancini, J.R. Marinelli, D.P. Menezes, M.M.W. Moraes, A.S. Schneider, Phys. Rev. C 76, 064318 (2007).

\bibitem{GRT} Y. K. Gambhir, P. Ring and A. Thimet,  Ann. Phys. {\bf 198} 132 (1990).

\bibitem{noble80} J.V.Noble, Phys. Lett. 89B, 325 (1980).

\bibitem{jennings91} M. Chiapparini, A. O. Gattone, and B. K. Jennings, Nucl.Phys. A529, 589 (1991).

\bibitem{Gal16} Gal, A., Hungerford, E.~V., \& Millener, D.~J.\ 2016, Reviews of Modern Physics, 88, 035004

\bibitem{ruester06} S.~B. R{\"u}ster, M. Hempel and J. Schaffner-Bielich, Phys. Rev. C {\bf 73}, 035804 (2006).

\bibitem{GP0} F. Grill, C. Provid{\^e}ncia, and S.~S.  Avancini, Phys. Rev. C {\bf 85}, 055808 (2012).

\bibitem{GP} F. Grill, H. Pais, C. Provid{\^e}ncia, I. Vida{\~n}a, and S.~S. Avancini, Phys. Rev. C {\bf 90}, 045803 (2014).

\bibitem{Sedrakian13} G.  Colucci and A. Sedrakian, Phys. Rev. C 87, 055806 (2013).

\bibitem{tov} J. R. Oppenheimer and G. M. Volkoff, Phys. Rev. {\bf 55}, 374 (1939); R. C. Tolman, {\it ibid.} {\bf 55}, 364 (1939).

%


\end{thebibliography}
\end{document}